\documentclass[apj]{emulateapj}

\usepackage{amsmath}
\usepackage{natbib}
\usepackage{hyperref}
\usepackage{graphicx}
\usepackage{epsf}
\usepackage{color}
\usepackage{threeparttable}
\usepackage{comment}
\usepackage{epsfig}
\usepackage{xspace}
\usepackage{bm}
\usepackage{subfigure}
\usepackage{ulem} 
\usepackage[usenames,dvipsnames,svgnames]{xcolor}
\usepackage{tabularx,ragged2e}
\newcolumntype{x}{>{\Centering}X}
\DeclareGraphicsExtensions{.jpg,.pdf,.png,.eps,.ps}

\newcommand{\lcdm}{\ensuremath{\Lambda\mathrm{CDM}}}

 \renewcommand{\vec}[1]{\mathbf{#1}}
 \newcommand{\Lk}{\mathcal{L}}

 \newcommand{\wmap}{\textit{WMAP}}
 
 \newcommand{\wnine}{\textit{WMAP}9}
 
  \newcommand{\sptsz}{SPT-SZ}
 \newcommand{\planck}{\textit{Planck}}

\def\hii{H\thinspace{$\scriptstyle{\rm II}$}~}

\def\be{\begin{equation}}
\def\ee{\end{equation}}
\def\ba{\begin{eqnarray}}
\def\ea{\end{eqnarray}}

\begin{document}
\title{A Comparison of Maps and Power Spectra 
Determined from South Pole Telescope and {\it Planck} Data}

\def\KICPChicago{1}
\def\AAUChicago{2}
\def\Davis{3}
\def\FNAL{4}
\def\ArgonneHEP{5}
\def\PhysicsUChicago{6}
\def\EFIChicago{7}
\def\SLAC{8}
\def\McGill{9}
\def\Caltech{10}
\def\Berkeley{11}
\def\Cifar{12}
\def\Colorado{13}
\def\ESO{14}
\def\Colphys{15}
\def\Illast{16}
\def\Illphys{17}
\def\UChicago{18}
\def\KIPAC{19}
\def\LBNL{20}
\def\Arizona{21}
\def\Michigan{22}
\def\Munich{23}
\def\ExcellenceCluster{24}
\def\MPE{25}
\def\Dunlap{26}
\def\Minnesota{27}
\def\Melbourne{28}
\def\CaseWestern{29}
\def\ArtInstChicago{30}
\def\JPL{31}
\def\CfA{32}
\def\Stanford{33}
\def\UToronto{34}

\author{
  Z.~Hou\altaffilmark{\KICPChicago,\AAUChicago},
  K.~Aylor\altaffilmark{\Davis},
  B.~A.~Benson\altaffilmark{\FNAL,\KICPChicago,\AAUChicago},
  L.~E.~Bleem\altaffilmark{\ArgonneHEP,\KICPChicago},
  J.~E.~Carlstrom\altaffilmark{\KICPChicago,\PhysicsUChicago,\ArgonneHEP,\AAUChicago,\EFIChicago},
  C.~L.~Chang\altaffilmark{\ArgonneHEP,\KICPChicago,\AAUChicago},
  H-M.~Cho\altaffilmark{\SLAC},
  R.~Chown\altaffilmark{\McGill},
  T.~M.~Crawford\altaffilmark{\KICPChicago,\AAUChicago},
  A.~T.~Crites\altaffilmark{\KICPChicago,\AAUChicago,\Caltech},
  T.~de~Haan\altaffilmark{\McGill,\Berkeley},
  M.~A.~Dobbs\altaffilmark{\McGill,\Cifar},
  W.~B.~Everett\altaffilmark{\Colorado},
  B.~Follin\altaffilmark{\Davis},
  E.~M.~George\altaffilmark{\Berkeley,\ESO},
  N.~W.~Halverson\altaffilmark{\Colorado,\Colphys},
  N.~L.~Harrington\altaffilmark{\Berkeley},
  G.~P.~Holder\altaffilmark{\McGill,\Cifar,\Illast,\Illphys},
  W.~L.~Holzapfel\altaffilmark{\Berkeley},
  J.~D.~Hrubes\altaffilmark{\UChicago},
  R.~Keisler\altaffilmark{\KICPChicago,\PhysicsUChicago,\KIPAC},
  L.~Knox\altaffilmark{\Davis},
  A.~T.~Lee\altaffilmark{\Berkeley,\LBNL},
  E.~M.~Leitch\altaffilmark{\KICPChicago,\AAUChicago},
  D.~Luong-Van\altaffilmark{\UChicago},
  D.~P.~Marrone\altaffilmark{\Arizona},
  J.~J.~McMahon\altaffilmark{\Michigan},
  S.~S.~Meyer\altaffilmark{\KICPChicago,\AAUChicago,\EFIChicago,\PhysicsUChicago},
  M.~Millea\altaffilmark{\Davis},
  L.~M.~Mocanu\altaffilmark{\KICPChicago,\AAUChicago},
  J.~J.~Mohr\altaffilmark{\Munich,\ExcellenceCluster,\MPE},
  T.~Natoli\altaffilmark{\KICPChicago,\PhysicsUChicago,\Dunlap},
  Y.~Omori\altaffilmark{\McGill},
  S.~Padin\altaffilmark{\KICPChicago,\AAUChicago},
  C.~Pryke\altaffilmark{\Minnesota},
  C.~L.~Reichardt\altaffilmark{\Berkeley,\Melbourne},
  J.~E.~Ruhl\altaffilmark{\CaseWestern},
  J.~T.~Sayre\altaffilmark{\CaseWestern,\Colorado},
  K.~K.~Schaffer\altaffilmark{\KICPChicago,\EFIChicago,\ArtInstChicago},
  E.~Shirokoff\altaffilmark{\Berkeley,\KICPChicago,\AAUChicago}, 
  Z.~Staniszewski\altaffilmark{\CaseWestern,\JPL},
  A.~A.~Stark\altaffilmark{\CfA},
  K.~T.~Story\altaffilmark{\KICPChicago,\PhysicsUChicago,\KIPAC,\Stanford},
  K.~Vanderlinde\altaffilmark{\Dunlap,\UToronto},
  J.~D.~Vieira\altaffilmark{\Illast,\Illphys}, and
  R.~Williamson\altaffilmark{\KICPChicago,\AAUChicago}
  }

\altaffiltext{\KICPChicago}{Kavli Institute for Cosmological Physics, University of Chicago, Chicago, IL, USA 60637}
\altaffiltext{\AAUChicago}{Department of Astronomy and Astrophysics, University of Chicago, Chicago, IL, USA 60637}
\altaffiltext{\Davis}{Department of Physics, University of California, Davis, CA, USA 95616}
\altaffiltext{\FNAL}{Fermi National Accelerator Laboratory, MS209, P.O. Box 500, Batavia, IL 60510}
\altaffiltext{\ArgonneHEP}{High Energy Physics Division, Argonne National Laboratory, Argonne, IL, USA 60439}
\altaffiltext{\PhysicsUChicago}{Department of Physics, University of Chicago, Chicago, IL, USA 60637}
\altaffiltext{\EFIChicago}{Enrico Fermi Institute, University of Chicago, Chicago, IL, USA 60637}
\altaffiltext{\SLAC}{SLAC National Accelerator Laboratory, 2575 Sand Hill Road, Menlo Park, CA 94025}
\altaffiltext{\McGill}{Department of Physics and McGill Space Institute, McGill University, Montreal, Quebec H3A 2T8, Canada}
\altaffiltext{\Caltech}{California Institute of Technology, Pasadena, CA, USA 91125}
\altaffiltext{\Berkeley}{Department of Physics, University of California, Berkeley, CA, USA 94720}
\altaffiltext{\Cifar}{Canadian Institute for Advanced Research, CIFAR Program in Cosmology and Gravity, Toronto, ON, M5G 1Z8, Canada}
\altaffiltext{\Colorado}{Center for Astrophysics and Space Astronomy, Department of Astrophysical and Planetary Sciences, University of Colorado, Boulder, CO, 80309}
\altaffiltext{\ESO}{European Southern Observatory, Karl-Schwarzschild-Stra{\ss}e 2, 85748 Garching, Germany}
\altaffiltext{\Colphys}{Department of Physics, University of Colorado, Boulder, CO, 80309}
\altaffiltext{\Illast}{Astronomy Department, University of Illinois at Urbana-Champaign, 1002 W. Green Street, Urbana, IL 61801, USA}
\altaffiltext{\Illphys}{Department of Physics, University of Illinois Urbana-Champaign, 1110 W. Green Street, Urbana, IL 61801, USA}
\altaffiltext{\UChicago}{University of Chicago, Chicago, IL, USA 60637}
\altaffiltext{\KIPAC}{Kavli Institute for Particle Astrophysics and Cosmology, Stanford University, 452 Lomita Mall, Stanford, CA 94305}
\altaffiltext{\LBNL}{Physics Division, Lawrence Berkeley National Laboratory, Berkeley, CA, USA 94720}
\altaffiltext{\Arizona}{Steward Observatory, University of Arizona, 933 North Cherry Avenue, Tucson, AZ 85721}
\altaffiltext{\Michigan}{Department of Physics, University of Michigan, Ann  Arbor, MI, USA 48109}
\altaffiltext{\Munich}{Faculty of Physics, Ludwig-Maximilians-Universit\"{a}t, 81679 M\"{u}nchen, Germany}
\altaffiltext{\ExcellenceCluster}{Excellence Cluster Universe, 85748 Garching, Germany}
\altaffiltext{\MPE}{Max-Planck-Institut f\"{u}r extraterrestrische Physik, 85748 Garching, Germany}
\altaffiltext{\Dunlap}{Dunlap Institute for Astronomy \& Astrophysics, University of Toronto, 50 St George St, Toronto, ON, M5S 3H4, Canada}
\altaffiltext{\Minnesota}{Department of Physics, University of Minnesota, Minneapolis, MN, USA 55455}
\altaffiltext{\Melbourne}{School of Physics, University of Melbourne, Parkville, VIC 3010, Australia}
\altaffiltext{\CaseWestern}{Physics Department, Center for Education and Research in Cosmology and Astrophysics, Case Western Reserve University,Cleveland, OH, USA 44106}
\altaffiltext{\ArtInstChicago}{Liberal Arts Department, School of the Art Institute of Chicago, Chicago, IL, USA 60603}
\altaffiltext{\JPL}{Jet Propulsion Laboratory, California Institute of Technology, Pasadena, CA 91109, USA}
\altaffiltext{\CfA}{Harvard-Smithsonian Center for Astrophysics, Cambridge, MA, USA 02138}
\altaffiltext{\Stanford}{Dept. of Physics, Stanford University, 382 Via Pueblo Mall, Stanford, CA 94305}
\altaffiltext{\UToronto}{Department of Astronomy \& Astrophysics, University of Toronto, 50 St George St, Toronto, ON, M5S 3H4, Canada}

\email{hou@kicp.uchicago.edu, kmaylor@ucdavis.edu}

\slugcomment{2018ApJ...853....3H}

\begin{abstract}
We study the consistency of 150 GHz data from the South Pole Telescope
(SPT) and 143 GHz data from the \planck\ satellite over the
patch of sky covered by the \sptsz{} survey.  We first
visually compare the maps and find that the residuals appear
consistent with noise after accounting for differences in angular
resolution and filtering.  We then calculate (1) the cross-spectrum
between two independent halves of SPT data, (2) the cross-spectrum
between two independent halves of \planck\ data, and (3) the
cross-spectrum between SPT and \planck\ data.  We find the three
cross-spectra are well-fit (PTE = 0.30) by the null hypothesis in
which both experiments have measured the same sky map up to a single
free calibration parameter---i.e., we find no evidence for systematic
errors in either data set. As a by-product, we
improve the precision of the SPT calibration by nearly an order of
magnitude, from 2.6\% to 0.3\% in power.  Finally, we compare all
three cross-spectra to the full-sky \planck\ power spectrum and find
marginal evidence for differences between the power spectra from the
\sptsz{} footprint and the full sky. We model these
differences as a power law in spherical harmonic multipole number. The best-fit
value of this tilt is consistent among the three cross-spectra in the
\sptsz{} footprint, implying that the source of this tilt is a sample
variance fluctuation in the \sptsz{} region relative to the full sky.
The consistency of cosmological parameters
derived from these datasets is discussed in a companion paper.
\end{abstract}


\maketitle

\section{Introduction}

One of the most remarkable results of modern cosmology is that 
a simple six-parameter model, usually referred to as the 
Lambda Cold Dark Matter (\lcdm) model, can fit the full range of cosmological observations. 
With the precision of cosmological observables now reaching the level of a few percent, however, several small discrepancies in the inferred parameter values are attracting attention.

These discrepancies show up in three places. 
The first is in the inferred parameter values from the CMB compared to some observations of the local universe. 
For example, the amplitude of
local density fluctuations, $\sigma_8$,  that is measured from observations of large scale structure appears lower
than the $\sigma_8$ value implied by cosmic microwave background (CMB)
measurements \citep{planck15-24}. 
There is also some tension between direct
measurements of the Hubble parameter $H_0$ and the value derived
from CMB measurements \citep{planck13-16, planck15-13, riess11, riess16}. 

Second, there are mild (1-2$\sigma$) discrepancies between parameter values derived from observations of the CMB by different experiments. 
 In particular, 
the best-fit parameters of the \lcdm\ model given \planck\ satellite
measurements of CMB temperature and polarization power spectra 
\citep{planck15-11} 
are somewhat different from those derived from earlier CMB data, whether from the 
Wilkinson Microwave Anisotropy Probe (\wmap) satellite \citep{hinshaw13}, 
from the combination of \wmap\ data and data from the South Pole Telescope 
(SPT, \citealt{hou14}), or from \wmap\ + SPT
plus data from the Atacama Cosmology Telescope (ACT, \citealt{calabrese13}).

Finally, work has been done on the internal consistency of $\Lambda$CDM model parameter values from different subsets of the \planck{} data. 
\citet{addison16} recently
pointed out that the matter density inferred from \planck\ data at $\ell
< 1000$ is 2.5 $\sigma$ discrepant from that inferred from \planck\ data
at $\ell > 1000$. 
In response, \citet{planck16-51} show that after correcting for certain
approximations in the \citet{addison16} analysis, and taking into account
the fact that matter density had been singled out as the most
discrepant parameter,  the global discrepancy is only 1.6 $\sigma$. 

Taken together, these low-level discrepancies have 
led some to speculate that we are seeing evidence of a 
potential failure of the \lcdm\ model \citep{wyman14,battye14},
systematic errors in the analysis of the low-redshift probes \citep{efstathiou14}, 
or systematic errors in the \planck\ data \citep{planck16-51}.  
The analysis presented in this paper is motivated by the fact that
 it is the high-$\ell$ temperature data from \planck\ that are driving
the parameter shifts of interest \citep{addison16,planck16-51}.
In terms of the measurement uncertainties, data from the 2540-square-degree SPT-SZ survey yield constraints within a 
factor of 2 of the \planck\ constraints at $\ell \gtrsim 1700$ and better than \planck\ at  $\ell \gtrsim 2100$ 
\citep[][hereafter S13]{story13}. 
The SPT-SZ data are thus a logical choice for consistency checks of
the high-$\ell$ \planck\ data. 

In this work, we present the first comparison between \planck\ and SPT data over the same patch of sky. 
Checks have previously been performed at the power-spectrum and cosmological-parameter level, for instance in \citet{planck13-16} and \citet{planck15-13}, and the two data sets have been shown to be roughly consistent, but again with some low-level discrepancies.
The strength of power-spectrum and parameter comparisons using the full datasets are limited by the sample variance of the SPT data at lower $\ell$ and \planck\ noise at higher $\ell$. 
By limiting the comparison to CMB modes measured by both experiments, we can greatly reduce the sample variance to sharpen the consistency tests between the two data sets.

Here we compare the SPT-SZ data in the 150 GHz band and \planck\ full-mission data in the 143 GHz band, restricted to the SPT-SZ observing region.
We calculate angular cross-spectra
of the SPT and \planck\ maps ($150\times 143$), and, for comparison, the cross-spectrum of one half of 
SPT data with the other half ($150\times 150$), and the cross-spectrum of one 
half of \planck\ data with the other half ($143\times 143$). 
We calculate the difference between 
the SPT $150 \times 150$ spectrum and the other two, as well as the ratio 
of the SPT $150 \times 150$ spectrum to the others. 
Using simulated observations of mock skies
(including realistic noise for both experiments), we calculate the expected uncertainty
in these differences and ratios, and we use this to calculate the $\chi^2$ and 
probability to exceed this $\chi^2$ under the null hypothesis that there are no systematic biases between the experiments.
This investigation is similar to that performed by \citet{louis14} on ACT and \planck\ data
over 592 deg$^2$ of sky and two observing bands (143/148~GHz and 217/218~GHz), the 
conclusion of which was that ACT and \planck\ measured statistically consistent CMB
fluctuations over that patch of sky.

This paper is organized as follows. In Section 2, we discuss the 
SPT 150 GHz and \planck\ 143 GHz temperature maps in the 2540 $\text{deg}^2$ 
patch of sky that constitutes the whole of the SPT-SZ survey region. In Section 
3 we present how power spectra are calculated from the maps and the simulations
generated to de-bias the data and build the covariance matrix.
In Section 4 we compare the power spectra and use simulations to test the null hypothesis 
that the two experiments are measuring the same sky, and we discuss these results
in the context of comparisons between SPT and full-sky \planck\ data.
We present our conclusions in Section 5.

\section{Data}

The main goal of this work is to compare maps and power spectra from the SPT 150\,GHz and \planck{} 143\,GHz data sets within the $2540\deg^2$ SPT-SZ survey area.
As an overview of the two datasets, Figure~\ref{fig:hemisphere} shows a sky map of \planck\ HFI 143 GHz full-mission data,
with the SPT-SZ survey area outlined by a solid black curve. 
In this section, we discuss details of the map-making and instrumental characteristics of each experiment.

\subsection{SPT}
\label{subsec:spt}
\begin{figure}
\begin{center}
   \includegraphics[width=0.5\textwidth, trim=5.5cm 0cm 5cm 0cm, clip=true]{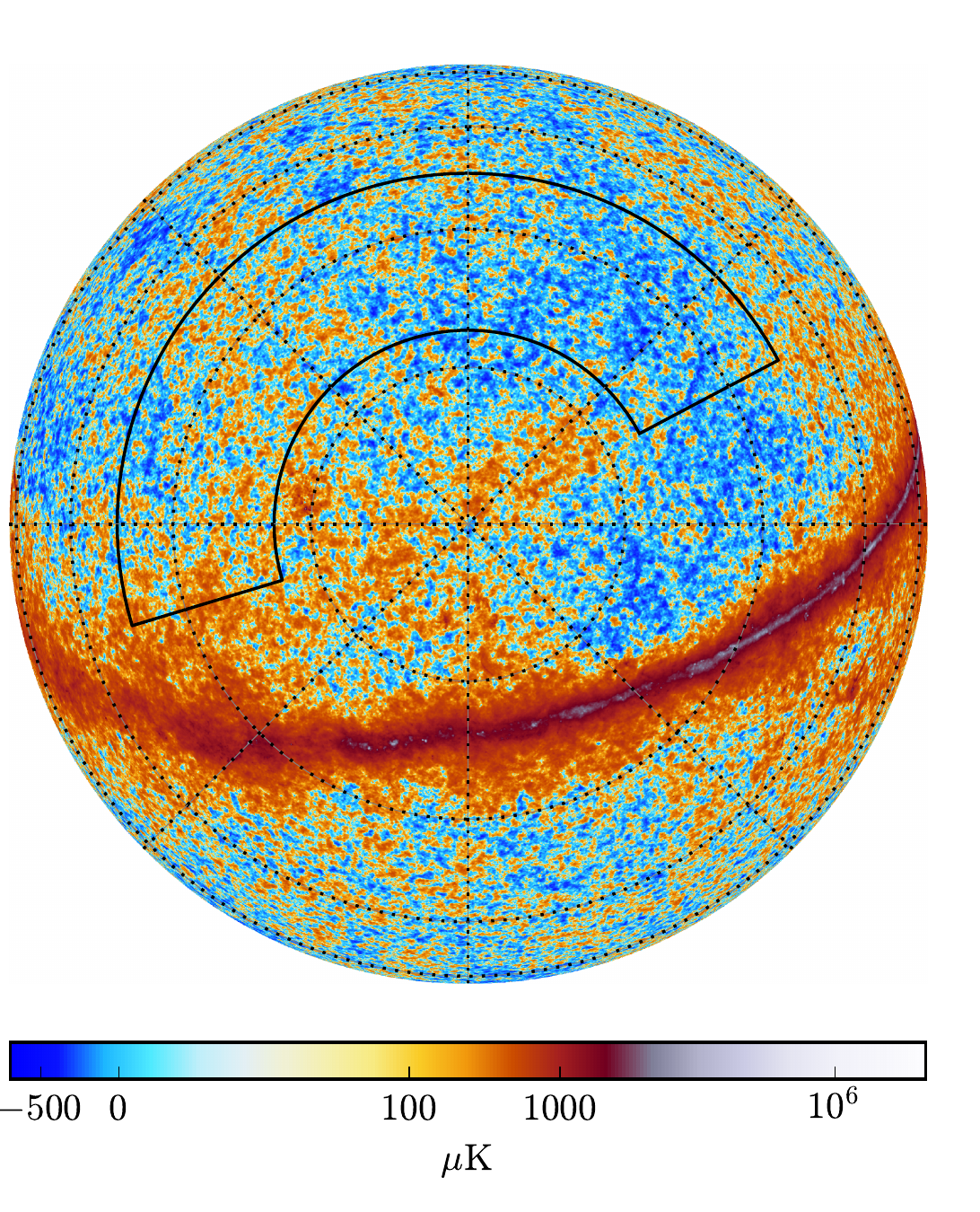}
\caption{The celestial southern hemisphere of CMB data from \planck\ HFI 143 GHz data.
The black curve 
outlines the SPT-SZ survey coverage. The image is oriented such that the line of $0^\mathrm{h}$ right ascension extends from the center to the top of the hemisphere, and right ascension increases counter-clockwise. 
}
\label{fig:hemisphere}
\end{center}
\end{figure}

The SPT is a 10-meter telescope located at the Amundsen-Scott South Pole station. From 2008 to 2011, the first camera on the SPT, a three-band bolometer array known as the SPT-SZ camera, was used to conduct a survey of $\sim$2500~deg$^2$ of the Southern sky with low Galactic dust contamination, referred to as the SPT-SZ survey. As shown in Figure~\ref{fig:hemisphere}, the survey area is a contiguous region extending from $20^\mathrm{h}$ to $7^\mathrm{h}$ in right ascension (R.A.) and from $-65^\circ$ to $-40^\circ$ in declination. 
The survey was conducted by observing a series of 19 areas, 
ranging in area from roughly 100 to 300 square degrees,
which together form the full survey region (see, e.g., S13, Figure 2).
If not specified, in this paper we use  ``field'' to refer to individual 
observing areas of SPT-SZ.

There were approximately 200 observations of each field, with each individual observation taking roughly 2 hours. Estimates of the primary CMB temperature power spectrum from this survey, and the resulting cosmological interpretation, are discussed in \citet[][hereafter K11]{keisler11}, S13, and \citet{hou14}. 
Our analysis is identical to that of S13 up to the single-observation map step. We briefly review 
that part of the analysis here and refer the reader to K11 and S13 for more details.

To process time-ordered data (TOD) into maps, the TOD from each SPT detector 
are first filtered and multiplied by a calibration factor. 
The filtering steps important for this analysis
are a high-pass filter and the subtraction of a common mode across all detectors in each of the six
160-element module in the focal plane. 
The high-pass filter cuts off signals below a temporal frequency corresponding to an angular 
frequency along the scan direction of $\ell=270$.
The common-mode subtraction acts as an isotropic high-pass filter with a cutoff at roughly
$\ell=500$. In both of these filtering steps, bright point sources ($S > 50$~mJy at 150~GHz)
are masked to avoid large filtering artifacts. 

The TOD from individual SPT detectors are then binned into maps using inverse-variance weighting, i.e.,
TOD samples corresponding to times when an individual detector was pointed at a particular pixel are averaged 
together using using inverse-variance weighting, then assigned to that pixel.
Maps are made using the oblique Lambert equal-area azimuthal projection \citep{snyder87} with 1-arcmin square pixels. 

The SPT maps  are simply binned-and-averaged maps of filtered data and thus are biased representations of the sky.
The signal in these maps is the true sky signal convolved with the instrument beam, the 
effect of the TOD filtering, and the effect of binning. Beams are discussed below, and the filter transfer function
is estimated through simulations, as discussed in Section~\ref{subsec:tf}.

We use the S13 estimates of the SPT 150~GHz beam transfer function $B_\ell$ and its uncertainty, 
and refer the reader to S13 for more details. 
Briefly, the main lobe is measured using bright point sources in the survey fields, while the sidelobes are measured using observations of Jupiter. Venus observations are used to stitch the two together. 
The main lobe of the beam is well-approximated by a 1.2-arcmin FWHM Gaussian. The beam uncertainty arises from several 
statistical and systematic effects, including residual atmospheric noise in the maps of Venus and Jupiter, and the weak 
dependence of $B_\ell$ on the choice of radius used to stitch the inner and outer beam maps.

The TOD from each detector in each observation is calibrated using the response to an 
internal thermal source, which is in turn tied to the brightness of the Galactic \hii region
RCW38. For details of this calibration, see \citet{schaffer11}. The full-depth maps are then 
compared to CMB satellite data to provide the overall absolute calibration. In K11 and S13, 
the power spectrum of the full-depth maps was compared to the full-sky \wmap\ estimate of the 
CMB power spectrum in the multipole range $650 \le \ell \le 1000$. For this work, we 
initially use the calibration determined in \citet{george15}, using the comparison of SPT
power spectra to the full-sky \planck\ power spectrum in the multipole range $670 \le \ell \le 1170$; 
however, the map-based comparison to \planck\ undertaken here ends up providing
a significantly more precise absolute calibration, as detailed in Section~\ref{subsec:bp_comparison}.

\subsection{\planck}
\label{subsec:planck}

The \planck\ satellite \citep{planck13-1} was launched in 2009 by the 
European Space Agency, with the goal of measuring the CMB temperature
and polarization anisotropy with significantly better sensitivity, 
angular resolution, and wavelength coverage than was achieved by its space predecessor, 
the \wmap\ mission \citep{bennett13}.
\planck\ mapped the full sky in nine bands, ranging in 
frequency from 30 to 857~GHz. 
In this work, we use HFI data from the 2015 data release.
Specifically, we use the 143~GHz full-mission, halfmission-1, 
and halfmission-2 maps downloaded from the NASA/IPAC Infrared Science Archive 
(IRSA).\footnote{\url{http://irsa.ipac.caltech.edu/data/Planck/release_2/all-sky-maps/}}

In this work, we use a cross-spectrum pipeline similar to that used in K11 and S13 to compare SPT
and \planck\ data on the SPT-SZ sky patch. To use this pipeline for \planck\ data, 
the \planck\ maps must have the same pixelization, projection, and field definitions as the SPT maps.
To achieve this, we create mock TOD using the \planck\ 143 GHz full-mission or half-mission maps and the SPT pointing and detector weight information from individual SPT observations.  The mock TOD are then binned into maps in the same manner as the real SPT TOD, using the same projection and pixelization.
Instead of using the SPT bandpass filtering, a much simpler high-pass
filter is applied to the \planck\ mock TOD to simply remove the
signals on scales with $\ell<100$. Our simulations show that applying
this simple high-pass filter greatly improves the numerical stability
of our unbiased power spectrum calculations. In the power spectrum analysis described below, we use the projected \planck\ maps generated from the map-making pipeline with this simple high-pass filter applied to the \planck\ mock TOD.

The \planck\ 143~GHz beam has been measured using the \planck\ TOD around planets, such as Mars, Jupiter and Saturn \citep{planck13-7, planck15-7}. The beam window function $B(\ell)$ for individual frequency detector sets has been released by the \planck\ collaboration as part of the ``Reduced Instrument Model'' (\href{http://irsa.ipac.caltech.edu/data/Planck/release_2/ancillary-data/ }{RIMO}).\footnote{\url{http://irsa.ipac.caltech.edu/data/Planck/release_2/ancillary-data/ }}
From the 2013 \planck\ data realease to the 2015 data release, there was a marked improvement
in the beam characterization, such that the uncertainty on the 2015 beam window functions for 143~GHz data 
is at the 0.1\% level for the $\ell$ range of interest to this work. 

\section{Power Spectrum}
\label{sec:powspec}

We now turn to  the methodology for calculating unbiased cross-spectra of SPT 150 GHz maps and \planck\ 143 GHz maps. 
We introduce the power spectrum estimator in Section~\ref{subsec:cross-spectra}. 
Simulations are used in several places while estimating the power spectra; we describe how these simulations are generated in Section~\ref{subsec:sims}. 
We discuss how these simulations are used to estimate the SPT filter transfer function
in Section~\ref{subsec:tf} and to calculate the noise and sample-variance parts of the 
power spectrum covariance in Section~\ref{subsec:covariance}.
We describe how beam uncertainty is incorporated into the analysis in Section~\ref{subsec:beam_error}.
Finally, we discuss how to calculate the window functions needed to compare the binned cross-spectrum estimates---or ``bandpowers''---to a theory spectrum in Section~\ref{subsec:bp_wfcorr}.

\subsection{Power Spectrum Estimator}
\label{subsec:cross-spectra}

We use an estimator similar to that used in K11 and S13 to calculate the various cross-spectra in this work.
Each cross-spectrum is calculated by correlating maps made from different observations of the same field---either correlating full-depth SPT maps with full-depth \planck\ maps or correlating two half-depth maps from the same experiment. 
In the latter case, there is a noise penalty relative to K11 or S13 who used $\mathcal{O}(100)$ independent maps depending on the field. 
However, the noise penalty is largely insignificant for the SPT maps (less than 2\% in power) on the angular scales of interest, and 
it is unavoidable for the \planck{} maps because a larger number of independent splits are not publicly available. 
We mask and zero-pad each map before calculating its 2-dimensional Fourier transform, $\tilde{m}_{\pmb{\ell}}$.\footnote{When calculating angular power spectra in this work, we use the flat-sky approximation, in which we replace spherical harmonic transforms with two-dimensional Fourier transforms.} 
The mask is a product of an apodization window and a point-source mask.\footnote{In this work, we use the same point source masks used in S13 but slightly different apodization masks; we have confirmed using simulations that any effect of the different masking on power spectra or cosmological parameters is negligible for this analysis.
}
The maps and masks are zero-padded before Fourier-transforming such that each Fourier-space pixel has a width of $\delta_\ell = 5$. 
The raw bandpowers are the binned average of the cross-spectra between two maps $A$ and $B$ within a multipole bin $b$:
\begin{equation}
\hat{D}_b^{A\times B} = \langle \frac{\ell(\ell+1)}{2\pi} H_{\pmb{\ell}} \mathrm{Re} \left[ \tilde{m}^A_{\pmb{\ell}} \tilde{m}^{B*}_{\pmb{\ell}} \right]\rangle_{\ell \in b}.
\end{equation}
For the SPT-only power spectrum, $A$ and $B$ are the two SPT half-survey maps; for the SPT and \planck\ cross-spectrum, $A$ and $B$ are the full mission maps of each experiments; and for the \planck-only 143 power spectrum, $A$ and $B$ are the two \planck\ half-mission maps. 
$H_{\pmb{\ell}}$ in the above equation represents the 2-dimensional weighting of the Fourier modes that was used by S13 to handle the anisotropic noise in the SPT maps. 
While this weighting is suboptimal for \planck{} data, we still use
the same S13-derived weighting for all bandpowers  to minimize the differential sample variance between cross-spectra. 

Since the input maps are biased estimates of the true sky, due to effects as the application of 
a mask to the maps, the raw bandpowers $\hat{D}_b$ are a biased estimate of the true bandpowers, $D_b$.
 The biased and unbiased estimates are related by
\begin{equation}
\hat{D}_{b^\prime} \equiv K_{b^\prime b} D_b \, ,
\end{equation} 
 where 
 \begin{equation}
\label{eqn:kdef}
K_{bb^\prime}=P_{b\ell}\left(M_{\ell\ell^\prime}\,F_{\ell^\prime}B^{2}_{\ell^\prime}\right)Q_{\ell^\prime b^\prime},
\end{equation}
$Q_{\ell^\prime b^\prime}$ is the binning operator and $P_{b\ell}$ is its reciprocal, 
$M_{\ell\ell^\prime}$ is the mode-coupling matrix, 
$F_{\ell}$ is the filter transfer function which accounts for the signal suppressed by TOD filtering,
and $B_{\ell}$ is the beam function.
For details on the unbiasing procedure, see K11, S13, and \citet{hivon02}.

The unbiased bandpowers are calculated on a field-by-field basis and then combined. We combine the bandpowers obtained from individual fields using the effective area of single fields as the weighting, as in K11 and S13.

\subsection{Simulations}
\label{subsec:sims}

We use simulations both in the calculation of the bandpowers and to characterize the degree of consistency between the SPT and \planck{} cross-spectra over the \sptsz{} survey region. 
In this section, we turn our attention to how these simulations are created and used. 
The final product of this procedure is 400 sets of simulated unbiased bandpowers
for each of the three combinations of data ($150\times 150$, $150\times 143$, and $143\times 143$).

\subsubsection{Sky Signals}

 Our simulations include the following components: 1) a gravitationally lensed CMB signal, 2) thermal and kinematic Sunyaev-Zel'dovich (SZ) signals, 3) the cosmic infrared background (CIB) signal, and 4) emission from radio galaxies. 
 We generate 400 realizations of the lensed CMB with LensPix \citep{lewis05}, based on the best-fit  $\Lambda\mathrm{CDM}$ model from \planck\ 2015 $\mathrm{TT,TE,EE+lowP+lensing}$ \citep{planck15-13}. 
Modes are generated out to $\ell_{\mathrm{max}} = 8000$, well above the angular multipoles, $\ell < 2500$, used in this comparison. 
 The maps are stored in HEALPix \citep{gorski05} format with resolution parameter $N_{\mathrm{side}}=8192$. 
 
Unlike S13 which used Gaussian realizations for all extragalactic foregrounds, here we generate realizations of individual sources for the bright CIB galaxies ($6.4\,\mathrm{mJy} < S < 50\,\mathrm{mJy}$) and all radio galaxies (up to the flux cut at 50\,mJy). 
 The bright CIB galaxies are drawn from the modeled $\mathrm{d}N/\mathrm{d}S$ of  \citet{cai13}, while the radio galaxy $\mathrm{d}N/\mathrm{d}S$ is taken from \citet{dezotti05}. 
 In both cases, the amplitudes of the $\mathrm{d}N/\mathrm{d}S$  distribution are calibrated by the actual observations of \citet{mocanu13} at 150 GHz. 

The other extragalactic foregrounds, the thermal and kinematic SZ signals and the low-flux CIB, are treated as Gaussian. 
The shapes of the thermal and kinematic SZ angular power spectra are taken from the \citet{shaw10} and \citet{shaw12} models respectively, with the amplitudes set to the median values from \citet{george15}, $D_{3000}^{\rm tSZ} = 4.38\,\mathrm{\mu K}^2$ and $D_{3000}^{\rm kSZ} = 1.57\,\mathrm{\mu K}^2$. 
We similarly draw upon the median values and templates from  \citet{george15} for the CIB terms. 
The clustered CIB spectrum is taken to follow $D_\ell \propto \ell^{0.8}$ with an amplitude $D_{3000}^c = 3.46\,\mathrm{\mu K}^2$. 
The shot-noise or ``Poisson'' CIB power from galaxies dimmer than $6.4\,\mathrm{mJy}$ is taken to be $D_{3000}^{\rm P} = 9.16\,\mathrm{\mu K}^2$. 

Each component is scaled appropriately to the effective frequency of the SPT or \planck{} maps and then coadded together to create the final sky realization.

\subsubsection{\planck\ Noise Simulations}
\label{subsubsec:plancknoise}

The \planck\ collaboration released the 8th Full Focal Plane (FFP8) simulation in the 2015 data release \citep{planck15-12}. There are 1000 full-mission FFP8 noise simulations available at \href{https://crd.lbl.gov/departments/computational-science/c3/c3-research/cosmic-microwave-background/cmb-data-at-nersc/ }{NERSC}.\footnote{\url{https://crd.lbl.gov/departments/computational-science/c3/c3-research/cosmic-microwave-background/cmb-data-at-nersc/ }} We use the FFP8 full-mission simulations to create noisy \planck-like realizations to characterize the noise contribution to the SPT-\planck\ cross ($150\times 143$) bandpowers as discussed in Section~\ref{subsec:bp_comparison}.

\begin{figure}
    \includegraphics[width=0.48\textwidth, trim=0.8cm 0 0 3.5cm]{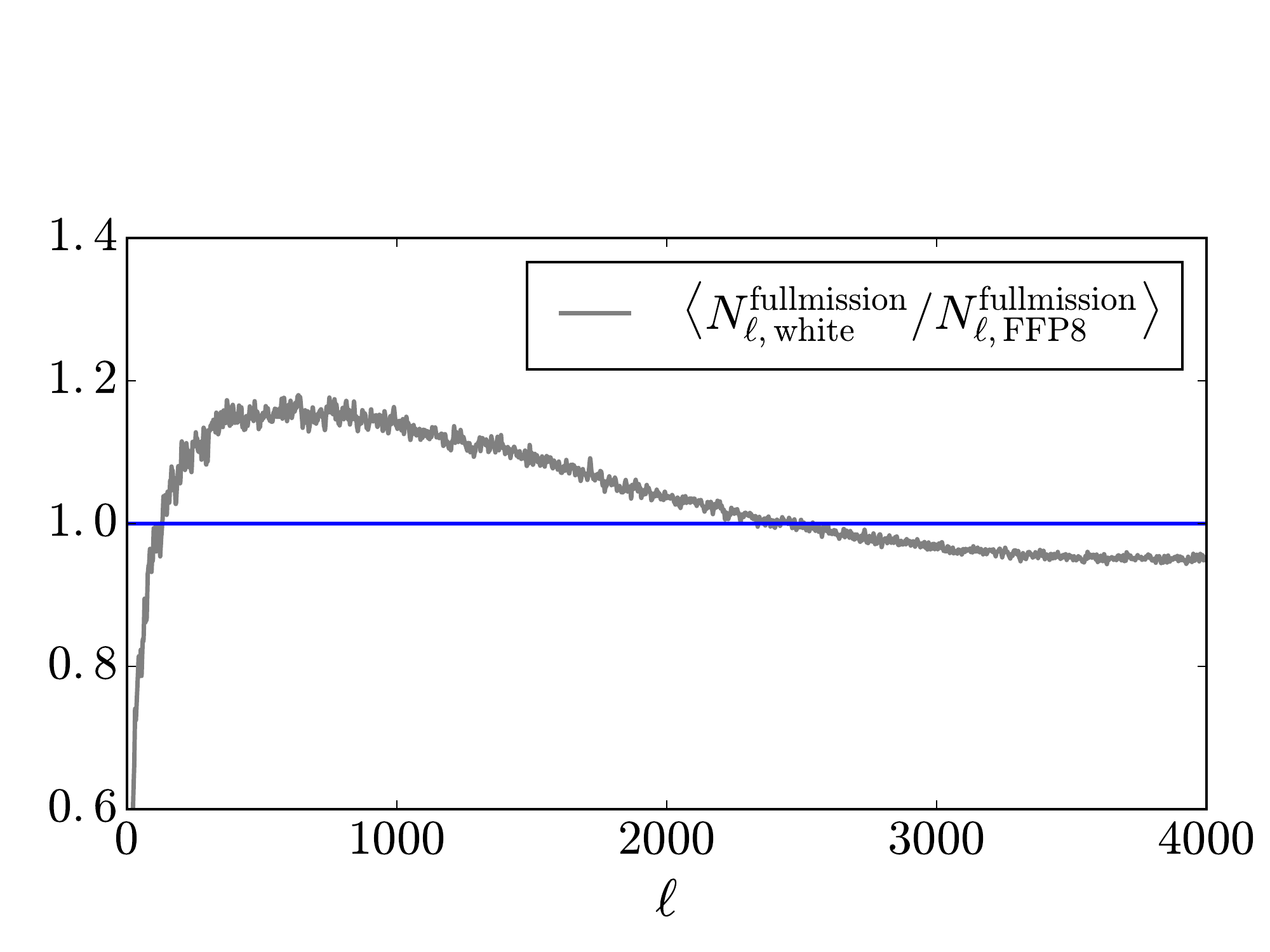}
\caption{Averaged ratio of the noise power spectrum between the white-noise realizations and \planck\ FFP8 noise simulations within the SPT survey area. 
While we use the full FFP8 noise simulations for the most-constraining set of bandpowers (SPT\,$\times$\,\planck), we use the simpler white-noise realizations when estimating uncertainties for the cross-spectrum between half-survey \planck{} maps.
}
\label{fig:cov_halfmission}
\end{figure}

The FFP8 noise simulations are only available for the full-mission data, and thus cannot be used for the \planck\ half-mission cross-spectrum. To create unbiased \planck\ 143 GHz bandpowers within the SPT area, we also generate half-mission noise simulations to characterize the noise contribution to the \planck-\planck\ ($143\times 143$) bandpowers as presented in Section~\ref{subsec:bp_comparison}. 
The half-mission noise simulations are based on the pixel-space noise variance released by the \planck\ collaboration along with the half-mission observation maps. 
We approximate the \planck\ map noise as Gaussian and uncorrelated between pixels (``white'') for the half-mission noise simulations.

To test this white-noise approximation, we compare the noise power spectrum over the SPT survey area from 100 of these Gaussian white-noise realizations to 100 FFP8 full-mission noise simulations. 
 The average power ratio is plotted in Figure~\ref{fig:cov_halfmission}.
 Clearly, the white-noise realizations overestimate the noise power  for $150<\ell<2200$, i.e. most angular multipoles of interest. 
 This implies that the \planck\ noise contribution is overestimated in our \planck-only bandpowers, 
 at a maximum level of roughly 15\% and a mean level of $\lesssim10\%$ across the $\ell$ range of interest.
We discuss the possible impact of this overestimate on our results in Section~\ref{subsec:bp_comparison}
and find that our main conclusions would likely be unchanged had we used more realistic noise simulations
for the \planck-only bandpowers.

\subsubsection{SPT Noise Realizations}\label{subsec:sptnoise}

SPT noise realizations are created directly from the data. For each individual SPT observation of each \sptsz{} survey field, we create a residual map which is the difference between a map made from all left-going telescope scans and a map made from all right-going telescope scans. 
There are approximately 200 jackknife maps for each field (the number of individual $\sim$2-hour observations). We use these maps to create the noise part of the simulated SPT observations. For each field, we multiply +1 or -1 randomly to the residual maps and coadd them together to form one noise realization. Using this method, we create 400 noise realizations.

With the same method, we also create SPT noise realizations of the first and second half sets of observations by coadding the residual maps of half of the observations within one field. Similar to the half-mission noise simulations for \planck\, we use the SPT half noise simulations to characterize the noise contribution to the SPT-only bandpowers.

\subsection{Filtering Transfer Function}
\label{subsec:tf}
The noise-free simulated observations are used to calculate $F_{\ell}$, the filtering transfer function, and $H_{\pmb{\ell}}$, the two-dimensional weight in the power-spectrum calculation. Given the input power spectrum of the simulations, the effective transfer function can be derived by comparing the power spectrum of the simulations with the known input spectrum using an iterative scheme \citep{hivon02}.
This method was used in K11 and S13; in this analysis, we make adjustments to Eq.6 and Eq.7 in \citet{keisler11} to include the \planck\ beam and pixel window function for the SPT-\planck\ cross-spectrum and \planck-only spectrum.

\subsection{Consistency Metrics}

We have three sets of unbiased bandpowers derived from SPT and \planck\ data, $\mathcal{D}_b^{150\times 150}$, $\mathcal{D}_b^{150\times 143}$, and $\mathcal{D}_b^{143\times 143}$. In this analysis, we use the difference between these bandpowers, in particular the residual and the ratio between the bandpower sets, to quantitatively characterize the consistency between the SPT and \planck\ datasets. 
This section presents the metrics we use for these consistency tests.

\subsubsection{Covariance Estimation}
\label{subsec:covariance}

We estimate the bandpower sample variance and noise variance using 400 sets of simulated signal+noise maps. 
The final bandpower covariance matrix will also include a contribution due to beam uncertainties; the calculation of the beam covariance is detailed in the next section. 
For a single field and data combination, the bandpower covariance, $\Xi^{X\times Y}$, 
is calculated simply as
\begin{equation}
\Xi^{X\times Y} = \left \langle \left ( \mathbf{D}^{X\times Y}_{b,\mathrm{sim}} - \mathbf{\bar{D}}^{X\times Y}_{b,\mathrm{sim}} \right ) \left ( \mathbf{D}^{X\times Y}_{b',\mathrm{sim}} - \mathbf{\bar{D}}^{X\times Y}_{b',\mathrm{sim}} \right ) \right \rangle,
\end{equation}
where $\mathbf{\bar{D}}^{X\times Y}_{b,\mathrm{sim}}$ is the mean over all 400 simulations for cross-bandpowers $X\times Y \in [150\times 150, 150 \times 143, 143\times143]$. 
The simulated $150\times 143$ cross-bandpowers $\mathcal{D}_{b,\mathrm{sim}}^{150\times 143}$ are derived from cross-spectra of simulated SPT maps including realizations of SPT full-observation noise and simulated \planck\ maps of the same underlying sky signal including \planck\ FFP8 noise simulations. 
For $\mathcal{D}_{b,\mathrm{sim}}^{150\times 150}$, the simulated bandpowers are calculated by the cross correlation of the two sets of simulated SPT maps with half-observation noise realizations. Similarly, the simulated $\mathcal{D}_{b,\mathrm{sim}}^{143\times 143}$ bandpowers are obtained using two sets of 
simulated \planck\ maps including \planck\ half-mission white-noise realizations.
The full  bandpower covariance matrix is then obtained by combining the individual-field covariance matrices with the square of the area weighting used to combine the bandpowers themseeves.

As we will evaluate consistency by looking at the differences and ratios among sets of bandpowers, 
we also need the covariance matrix for these quantities. 
For the differences or residual bandpowers, $\Delta\mathbf{D}_\mathrm{b} \equiv \mathbf{D}^{i \times j}_b - \mathbf{D}^{m \times n}_b$
(where $i,j,m,n \in \{150,143\}$), the  noise-plus-sample-variance
part of the covariance is easily estimated from the 400 simulations: 
\begin{equation}
\Xi_\mathrm{resid} = \langle \Delta\mathbf{D}_{b,\mathrm{sim}} \Delta\mathbf{D}_{b',\mathrm{sim}} \rangle.
\end{equation}
For the ratios relative to the $150\times 150$ bandpowers, the covariance can be expressed as:
\begin{equation}
\Xi_\mathrm{ratio} = \left \langle \left ( \frac{\mathbf{D}^{i\times j}_{b,\mathrm{sim}} }{ \mathbf{D}^{150 \times 150}_{b,\mathrm{sim}}} - 1 \right ) \left ( \frac{\mathbf{D}^{i\times j}_{b',\mathrm{sim}}}{ \mathbf{D}^{150 \times 150}_{b',\mathrm{sim}}} - 1 \right ) \right \rangle.
\end{equation}

\subsubsection{Beam Uncertainty}
\label{subsec:beam_error}

In this section, we present how beam uncertainties are handled for the bandpower comparison. 
Beam uncertainties appear as a second covariance term, in addition to the noise-plus-sample-variance term presented above, because the simulations did not include beam uncertainties. 

We begin by estimating the eigenmodes of the \planck{} and SPT beam covariance matrices. The \planck{} beam eigenmodes can be obtained from RIMO in 2015 data release, and the SPT beam eigenmodes can be derived from the beam 
correlation matrix calculated in S13.
For each eigenmode, we can calculate the fractional beam uncertainty as a function of multipole, $\delta b_{\ell}/b_\ell$, and propagate this linearly to the bandpower space according to: 
\begin{equation}
\delta D^{i\times j}_{b,\mathrm{beam}} = - \sum_{\ell}W^{i\times j}_{b\ell} D^{i\times j, \mathrm{fid}}_{\ell}\left(\frac{\delta b^i_\ell}{b^i_\ell} + \frac{\delta b^j_\ell}{b^j_\ell}\right), 
\end{equation}
where $i,j\in\{150,143\}$, $W^{i\times j}_{b\ell}$ is the bandpower window function, and $D_\ell^{i\times j, \mathrm{fid}}$ is the fiducial power spectrum  including CMB and extra-galactic foregrounds for that frequency combination. 

For the consistency tests, we need the beam covariance for bandpower differences and bandpower ratios. 
For the bandpower differences, the above equation leads straightforwardly to: 
\begin{eqnarray}
&& \Xi^{(i\times j)(m\times n)}_{bb',\mathrm{beam}}  = \\
\nonumber &&\sum_\mathrm{e-modes} (\delta D^{i \times j}_{b,\mathrm{beam}} - \delta D^{150\times 150}_{b,\mathrm{beam}}) (\delta D^{m \times n}_{b',\mathrm{beam}} - \delta D^{150\times 150}_{b',\mathrm{beam}}) ,
\label{eqn:covmat}
\end{eqnarray}
where $i,j,m,n \in \{150,143\}$ and the sum is taken over all eigenmodes of the beam covariance matrices. 

For the bandpower ratios, the beam covariance can be written as:
\begin{eqnarray}
&& \Xi^{(i\times j)(m\times n)}_{bb',\mathrm{beam}}  = \\
\nonumber &&\sum_\mathrm{e-modes} \left(\frac{\delta D^{i \times j}_{b,\mathrm{beam}} - \delta D^{150\times 150}_{b,\mathrm{beam}}}{D_b^{150\times 150}}\right) \left(\frac{\delta D^{m \times n}_{b',\mathrm{beam}} - \delta D^{150\times 150}_{b',\mathrm{beam}}}{D_{b'}^{150\times 150}}\right).
\label{eqn:covmat}
\end{eqnarray}

The total bandpower covariance for a consistency test is then the sum of the noise-plus-sample-variance term from the last section and the appropriate beam covariance term above.

\subsubsection{Bandpower Window Function Correction}
\label{subsec:bp_wfcorr}

\begin{figure}
\includegraphics[width=0.48\textwidth, trim=0 0 1cm 0cm]{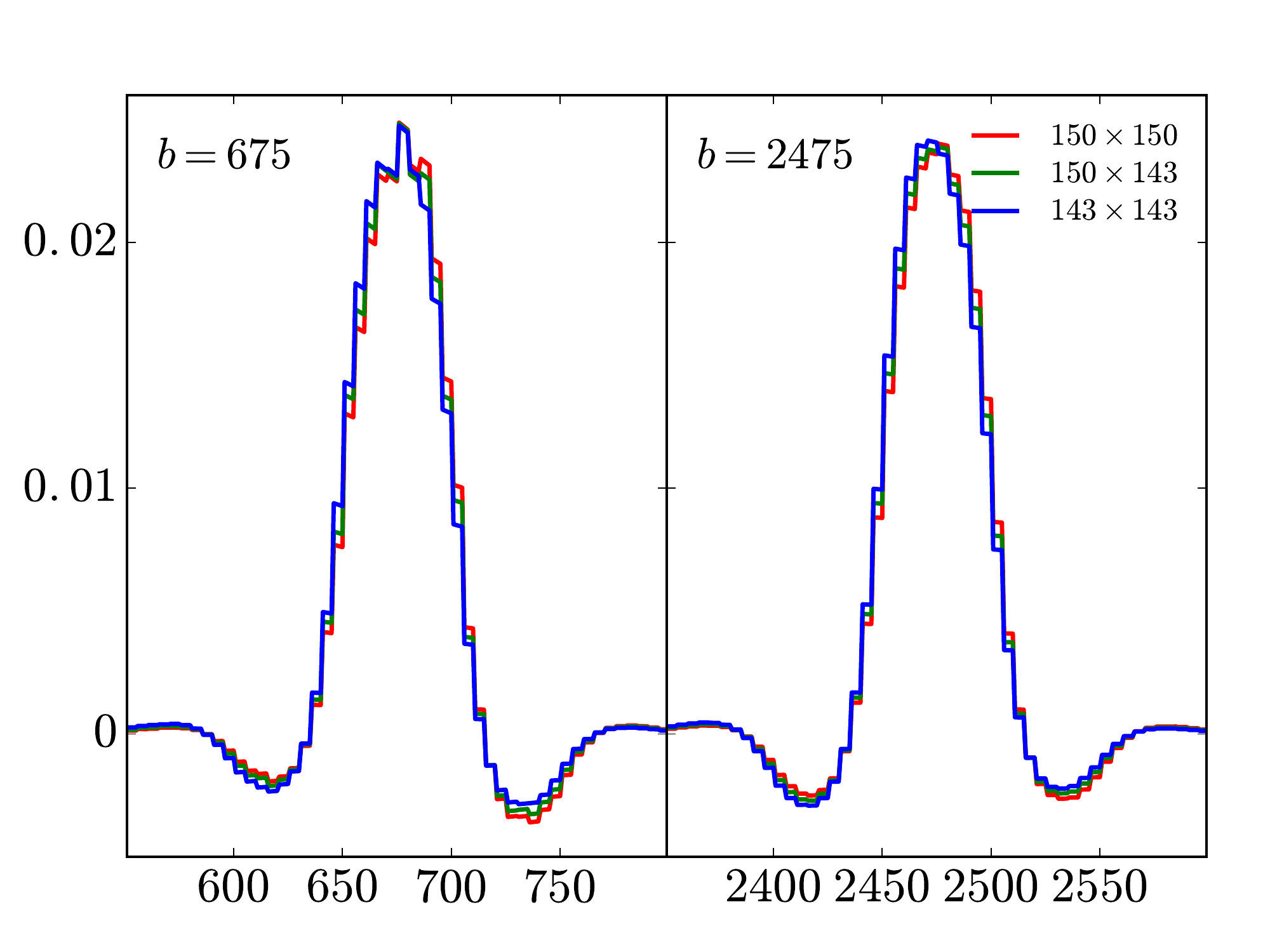}
\caption{The window functions $W_{b\ell}$ for the $D_b^{150\times 150}$ (red), $D_b^{150\times 143}$ (green), and $D_b^{143\times 143}$ (blue) bandpowers, for bins centered at $\ell=675$ and $2475$. In both panels, $W_{b\ell}^{143 \times 143}$ has a higher (lower) weight on $\ell$s lower (higher) than the bin center of $W_{b\ell}^{150 \times 150}$, and $W_{b\ell}^{150 \times 143}$ sits between the two. Due to the shape of CMB power spectrum, $D_\ell$, such difference leads to an $\ell$-dependent bias when directly comparing the three sets of bandpowers.}
\label{fig:winfunc}
\end{figure}

To characterize the level of consistency between the three sets of bandpowers ($150\times 150$, $150\times 143$, and $143\times 143$), we choose one set, $D_b^{150\times 150}$, as the fiducial which is subtracted from or divided into the other two sets (see Section~\ref{subsec:bp_comparison}). However, the residual bandpowers and bandpower ratios are biased due to the difference of bandpower window functions \citep[e.g.,][]{knox99}. Due to the different filtering and the beam transfer functions of the two experiments, the bandpower window functions are different between bandpower sets. In Figure~\ref{fig:winfunc}, the window functions of the three sets of bandpowers are illustrated for two bins with effective centers at $\ell = 675$ and $\ell = 2475$. 
In both of these bins, relative to the $150\times150$ window functions, the $143\times 143$ and $150\times 143$ bandpowers receive more weight from multipoles lower than the bin center and less from multipoles higher than the bin center.
These differences, if left unaddressed, lead to a bias in the bandpower differences that is dependent, to some degree, on the assumed cosmological model.

\begin{figure}
\includegraphics[width=0.48\textwidth, trim=1cm 0.5cm 0.5cm 0.5cm]{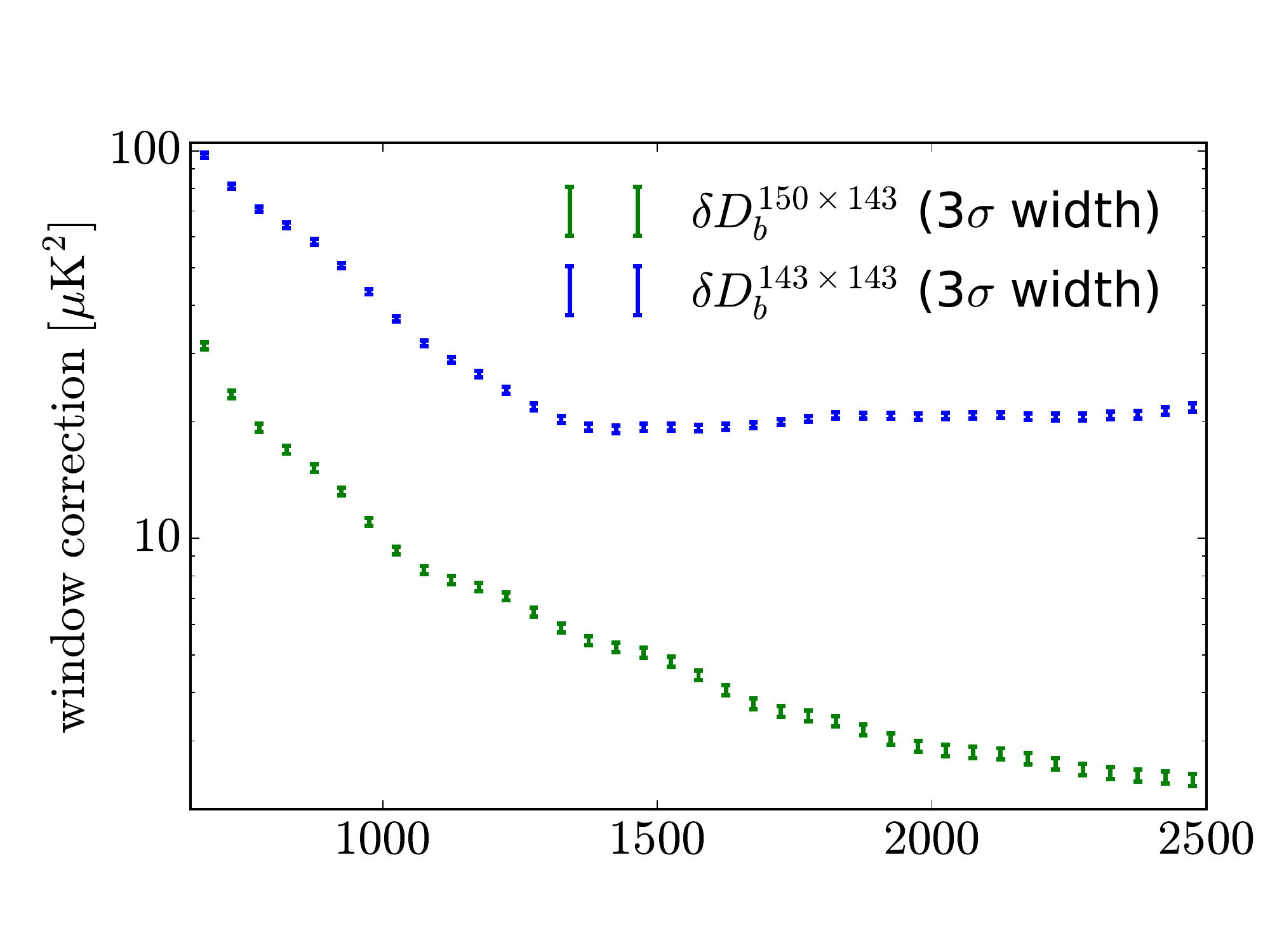}
\includegraphics[width=0.48\textwidth, trim=1cm 0.5cm 0.5cm 2.5cm]{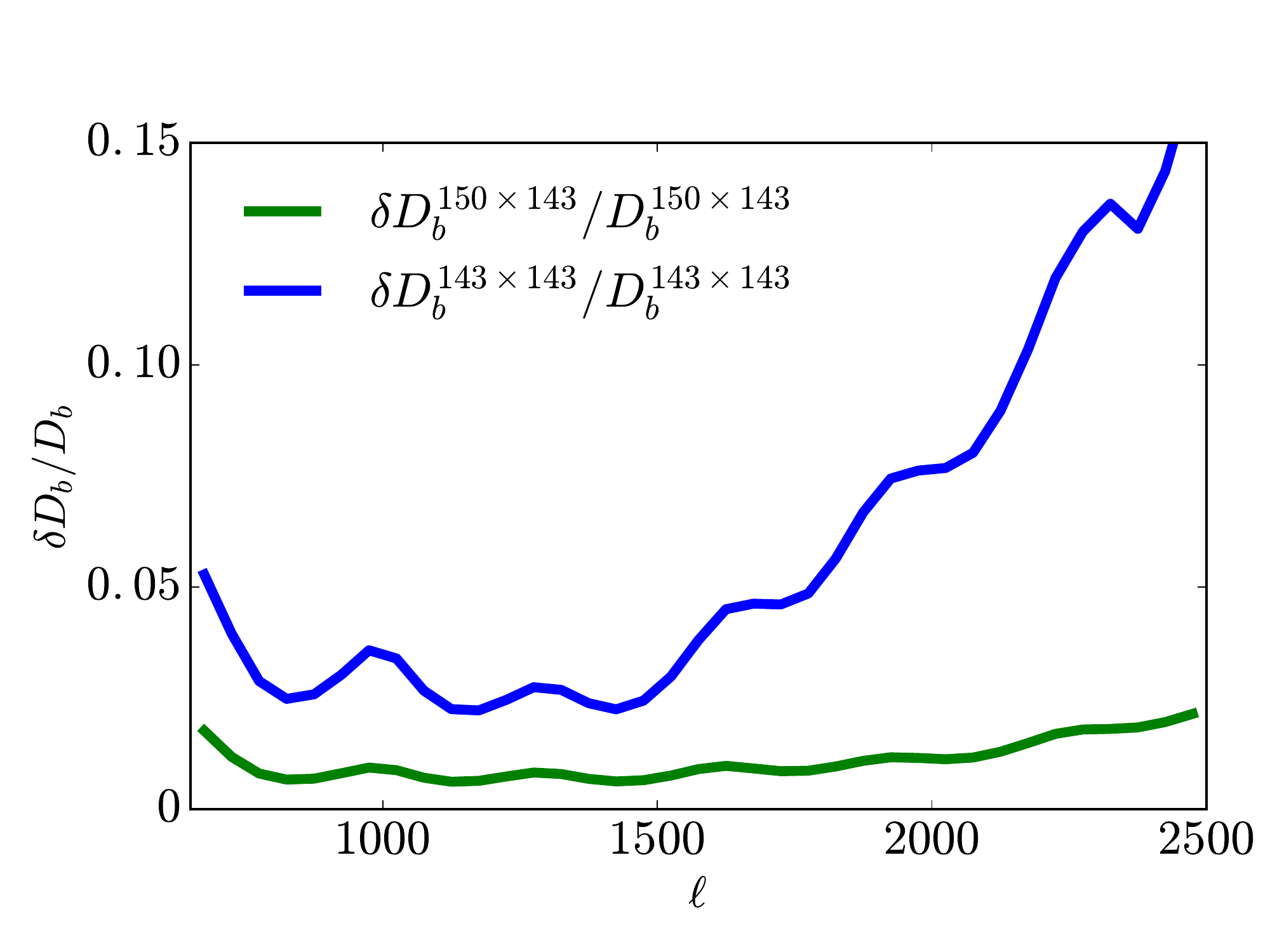}
\caption{\textit{Upper panel}: The bandpower window function correction as given by Eq.~\ref{eqn:bpwfunc_corr} for $150\times 143$ (green) and $143\times 143$ (blue), with the fiducial $D_\ell$ including the \planck\ 2015 best-fit cosmology and the best-fit foreground models. The error bars indicate the $3\sigma$ variation 
in the window function corrections marginalized over 
a Markov chain based on \wnine+SPT data. 
\textit{Lower panel}: The ratio between the window function correction $\delta D_b^i$ and the corresponding bandpower $D_b^i$ for $i= 150\times 143$ (green) and $143\times 143$ (blue).}
\label{fig:wfunc_corr}
\end{figure}

Before comparing the bandpower sets, we need to correct for this bias. The correction is calculated as follows 
\begin{equation}
    \delta D^i_b = \sum_\ell (W^i_\ell - W^{150\times 150}_\ell) D^{\rm fid}_\ell
\label{eqn:bpwfunc_corr}
\end{equation}
where $i = 150\times 143$, or $143\times 143$. 
The \planck\ 2015 best-fit cosmology plus the best-fit extra-galactic foregrounds are used as the fiducial model.
We subtract this correction term from the $150 \times 143$ and $143 \times 143$ bandpowers to remove the bias caused by the bandpower window function differences. In Figure~\ref{fig:wfunc_corr}, the upper panel shows the shape of the correction $\delta D_b$ and the lower panel shows the ratio between $\delta D_b$ and the corresponding $D_b$. For $150\times 143$, $\delta D_b$ yields a roughly flat $1\% - 2\%$ correction to the bandpower; while for $143\times 143$, the correction is more important at higher multipoles, up to $\sim 15\%$ at $\ell = 2500$. The window function correction is clearly critical for the consistency analysis between these three sets of bandpowers.

We now investigate the model dependence of the window function corrections. 
To characterize this dependence, we calculate the window function corrections for a distribution of input $D_\ell$. The distribution is obtained by randomly sampling the cosmology and foreground models from a Markov chain based on \wnine+SPT data. The variation of $\delta D_b$ in Eq.~\ref{eqn:bpwfunc_corr} is illustrated in the upper panel of Figure~\ref{fig:wfunc_corr} with the error bars indicating the $3\sigma$ variation of the corrections. Within the band range of interest, this variation level is less than 1\% of the bandpower standard error, we therefore ignore this very small uncertainty in our analysis.

\section{Results}
\label{sec:results}

In this section, we present the results of the comparison between SPT 150 GHz data and
\planck\ 143 GHz data. We first present a qualitative map-level comparison of the two data sets
in a common sky area. 
We then 
present the major result of the paper: the 
bandpower comparison among three sets of bandpowers calculated on the 2540 $\text{deg}^2$ 
SPT-SZ sky patch: $\mathcal{D}_b^{150\times 150}$, $\mathcal{D}_b^{150\times 143}$, and $\mathcal{D}_b^{143\times 143}$. Setting $\mathcal{D}_b^{150\times 150}$ as the fiducial and subtracting it from the other two, we use the residual as the quantity to characterize the level of consistency between SPT 150 GHz and \planck\ 143 GHz data. As a byproduct, for the first time we recalibrate the SPT data to the \planck\ data on the same patch of sky.

We also characterize the ratios between these sets of bandpowers. 
Note that the residual and ratio tests are not intended to be independent checks, as they contain
nearly the same information presented in different ways. The reason we use both metrics 
is that the systematic errors most likely to cause differences between SPT and \planck\ data 
(such as unmodeled foreground residuals or beam systematics) fall into two broad categories: 
additive or multiplicative systematics. Additive systematics will show up most obviously in the residual
bandpower test, while multiplicative systematics will show up most obviously in a bandpower ratio
test.

Finally, we compare each set of 
2540 $\text{deg}^2$ bandpowers to the full-sky \planck\ 143 GHz bandpowers. Obviously such bandpower comparisons are not distinct from the map comparison. Rather, the residuals and the ratios of these sets of bandpowers are the quantitative version of the by-eye map comparison.

\subsection{Map-level Comparison}
\label{subsec:mapcomp}
\begin{figure*}
\begin{center}
   \includegraphics[width=0.47\textwidth, trim=2.8cm 1.4cm 3.1cm
0.5cm]{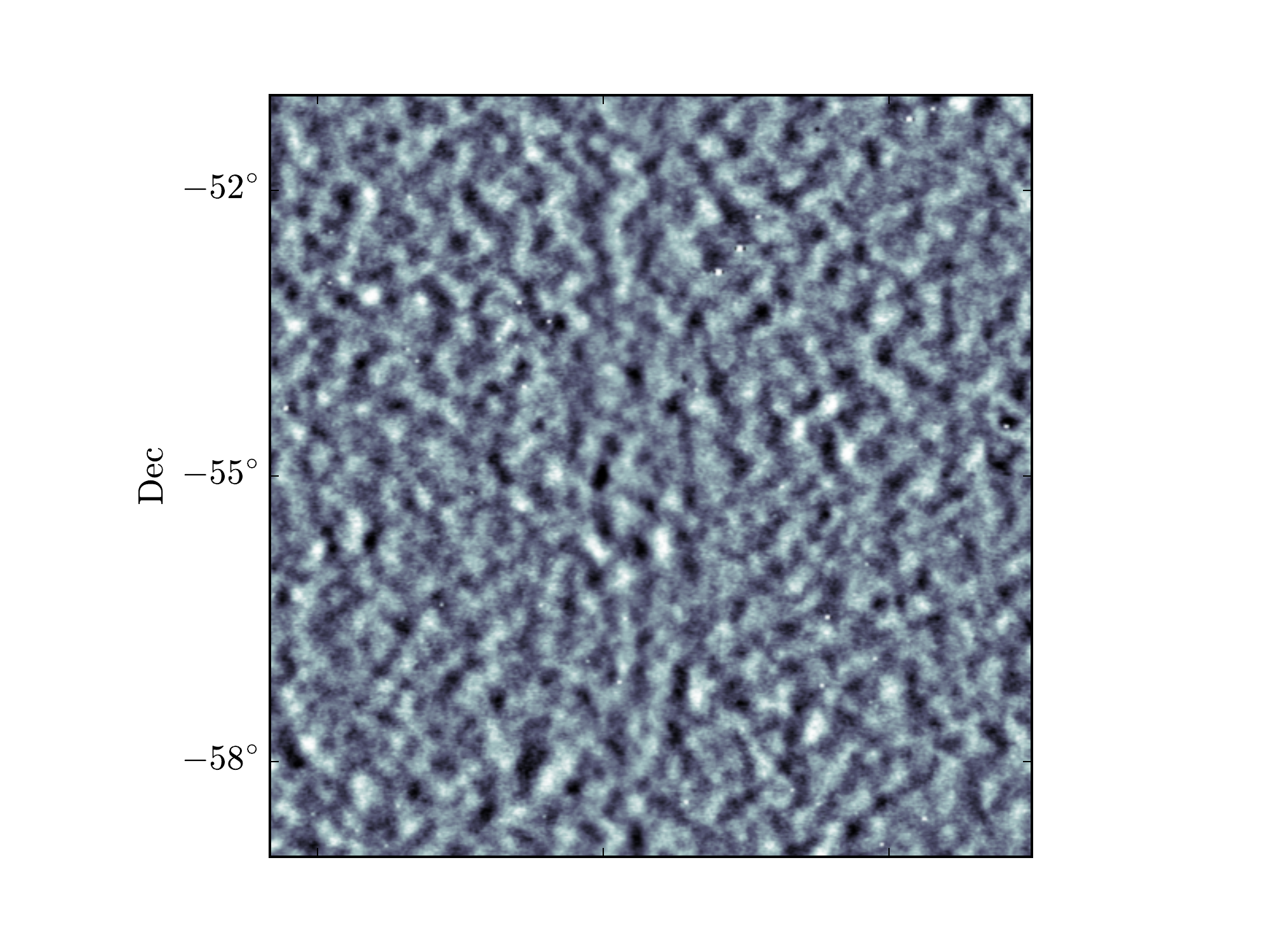}
   \includegraphics[width=0.47\textwidth, trim=4.3cm 1.4cm 1.6cm
0.5cm]{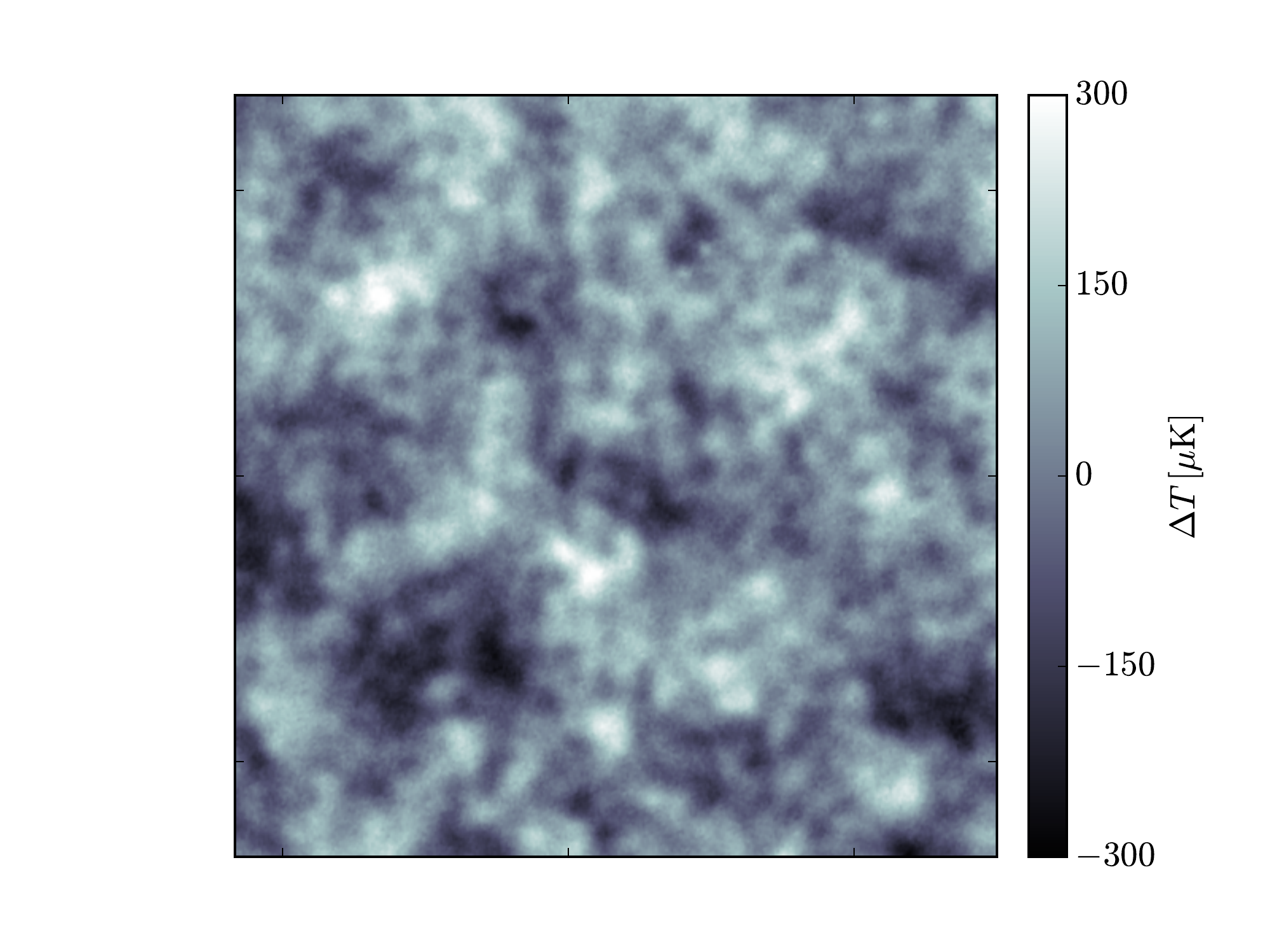}
    \includegraphics[width=0.47\textwidth, trim=2.8cm 0cm 3.1cm
1.4cm]{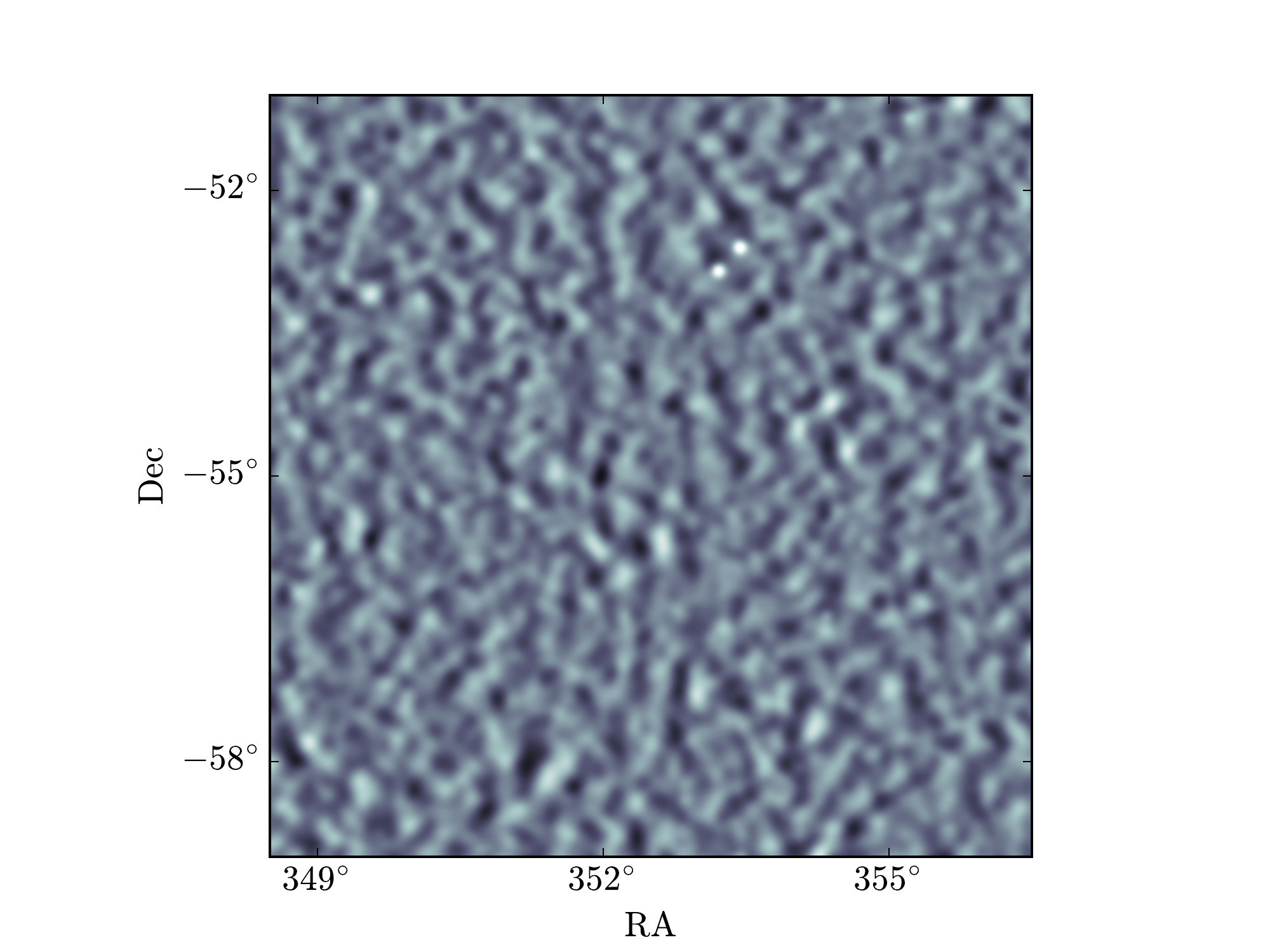}
    \includegraphics[width=0.47\textwidth, trim=4.3cm 0cm 1.6cm
1.4cm]{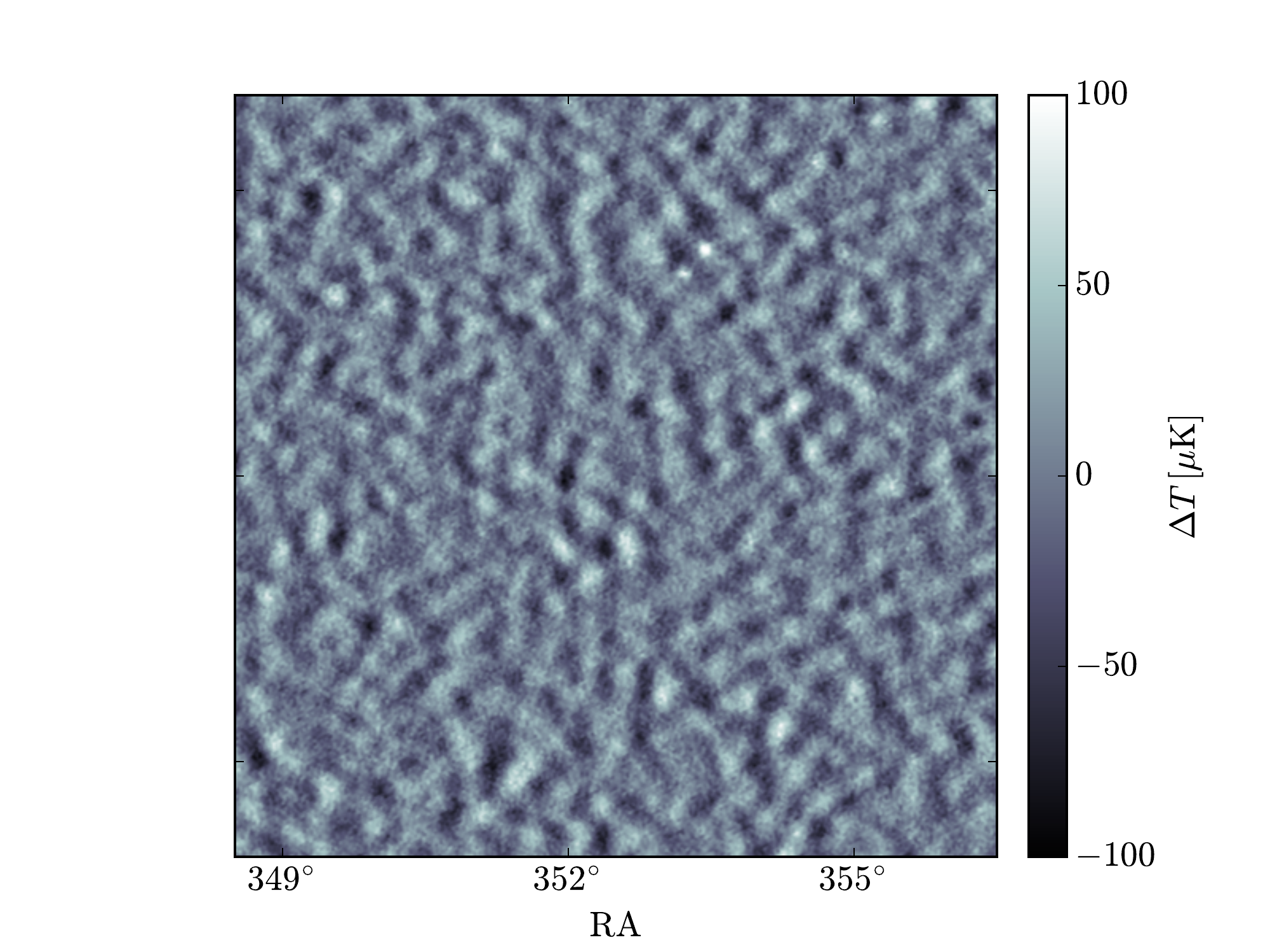}
\caption{Temperature maps of a portion of one SPT-SZ survey field. \textit{Upper left}: the SPT-SZ 150 GHz map of this region. Modes with $\ell\lesssim600$ are strongly suppressed in this map by the high-pass filter applied to the time-ordered data. \textit{Upper right}: the \planck\ 143 GHz full-mission map of the same area. \textit{Lower left}: the SPT map from the upper-left panel smoothed to have the same resolution as the \planck\ map. \textit{Lower right}: the \planck\ map from the upper-right panel with the SPT-SZ high-pass filter applied. 
The difference between the lower-left and lower-right panels is shown in Figure~\ref{fig:diffmap}.
\textbf{Note}: the grayscale range of the top right panel is $[-300\mathrm{\mu K},\mathrm{300 \mu K}]$, and the greyscale range of the other three panels is $[-100\mathrm{\mu K},\mathrm{100 \mu K}]$.}
\label{fig:map}
\end{center}
\end{figure*}

\begin{figure*}
\begin{center}
    \includegraphics[width=0.47\textwidth, trim=2.8cm 0cm 3.1cm
1.4cm]{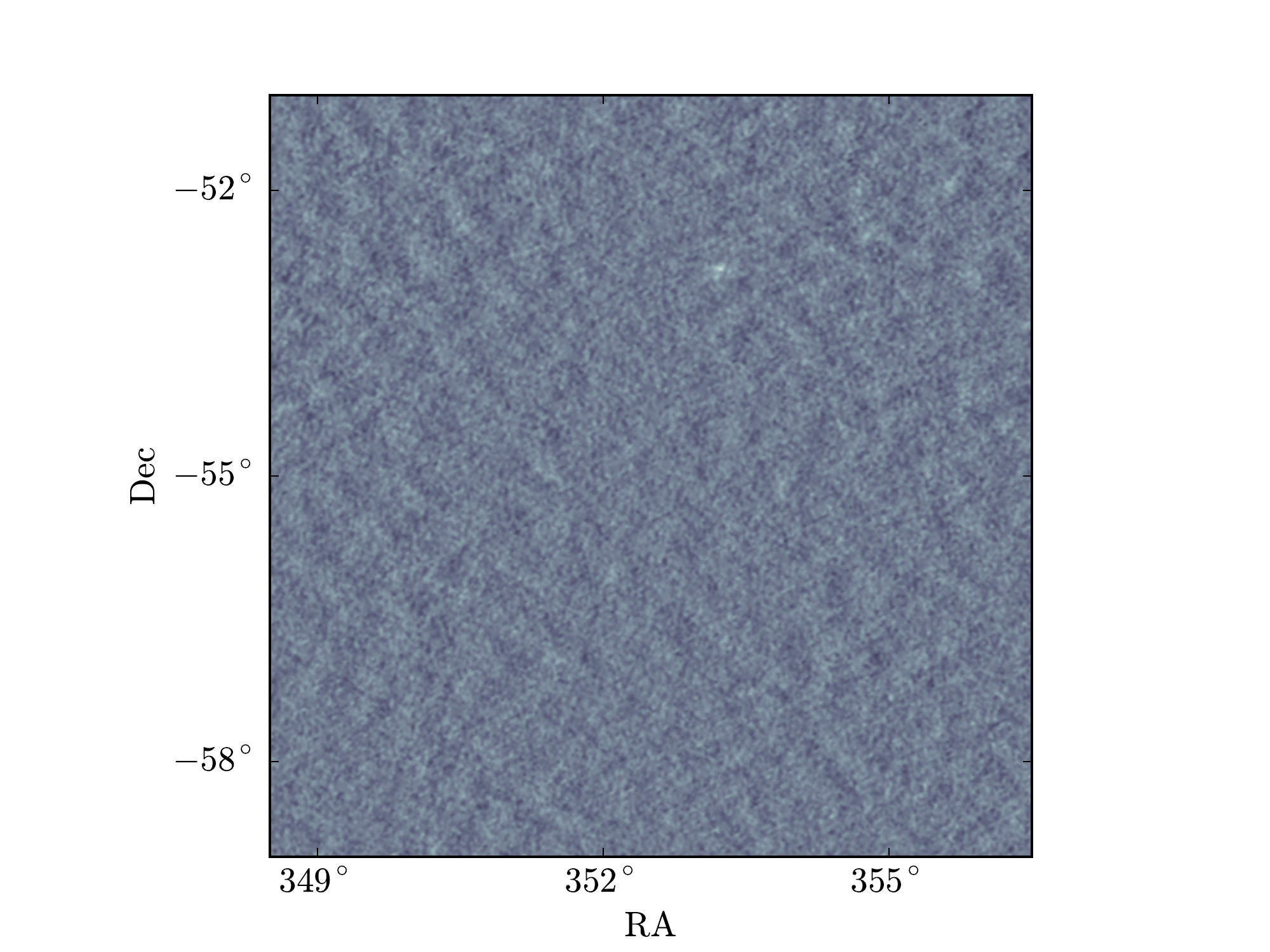}
    \includegraphics[width=0.47\textwidth, trim=4.3cm 0cm 1.6cm
1.4cm]{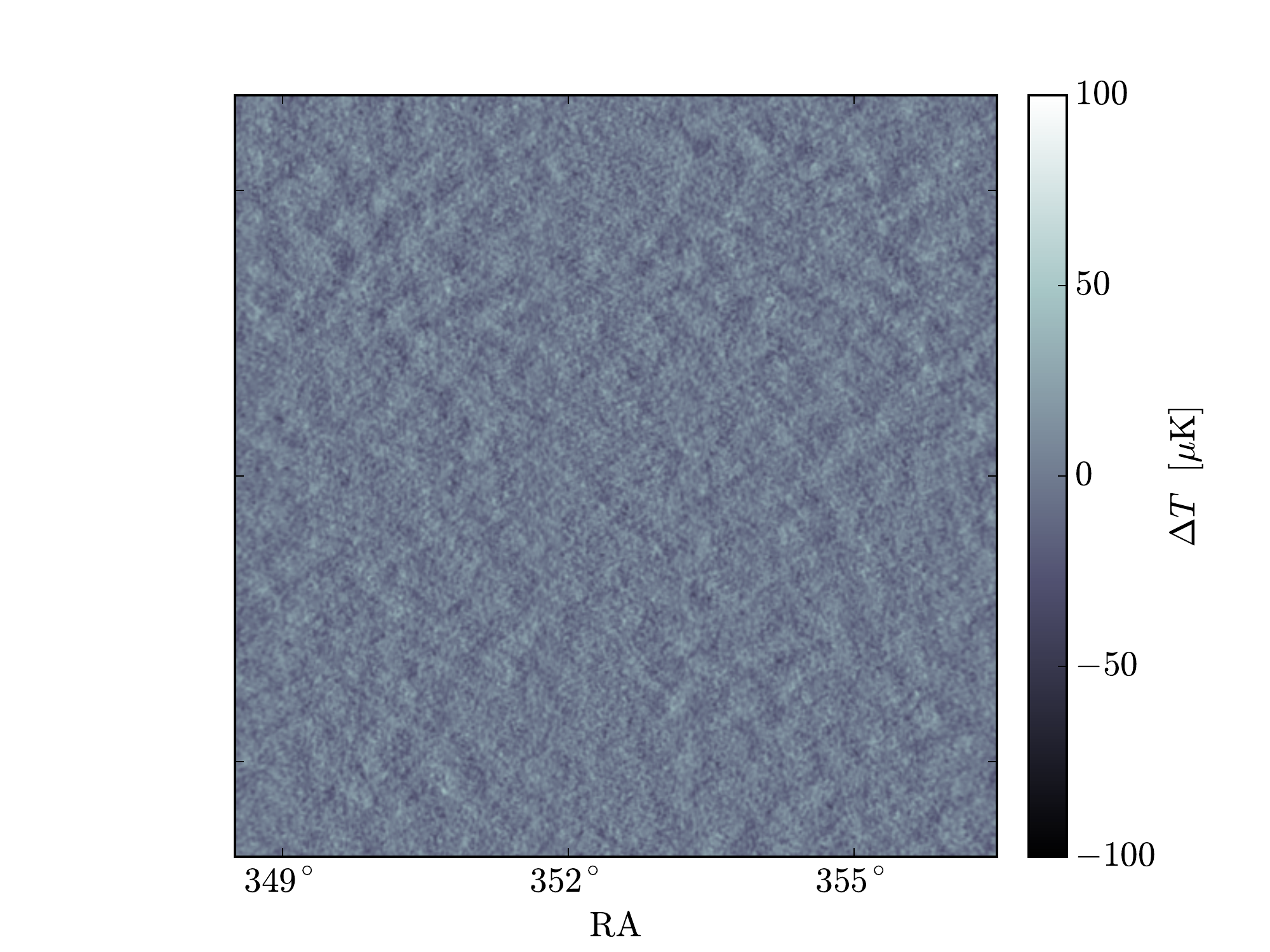}
\caption{
\textit{Left panel}: The residual map between the smoothed SPT-SZ 150 GHz map in the lower-left panel of 
Figure~\ref{fig:map} and the high-pass-filtered \planck\ 143 GHz map in the lower-right panel
of Figure~\ref{fig:map}. The feature at the location of one of the bright point sources is potentially due to 
temporal variability of the source, as the SPT and \planck\ data were not taken simultaneously.
\textit{Right panel}: Same as left panel, but with simulated data, consisting of the sum of an SPT-SZ 
noise realization smoothed to match \planck\ resolution and a \planck\ noise realization filtered to 
match the time-domain filtering of the SPT-SZ data.}
\label{fig:diffmap}
\end{center}
\end{figure*}

The SPT and \planck\ maps within the same patch are presented in Figure~\ref{fig:map} for visual comparison.
The upper panels show the filtered SPT map described in Section~\ref{subsec:spt} on the left and the projected \planck\ full-mission 143 GHz map on the right.
Some bright point sources can be identified by eye in both the SPT and \planck\ maps, but the maps do not resemble each other, because the \planck\ map is dominated by the degree-scale CMB anisotropy filtered out of the SPT data, and the SPT map shows more small-scale structure due to its higher angular resolution. 
In the lower panels of Figure~\ref{fig:map},
we show SPT and \planck\ maps of the same spatial modes. The lower-left panel shows the SPT map from the 
upper-left panel convolved with the difference between the SPT 150~GHz and \planck\ 143~GHz beams; 
the lower-right panel shows a map made by observing a \planck\ sky with the SPT 150~GHz scan strategy and TOD filtering.
Though the \planck\ map has a visibly higher noise level (as expected), the signals in the two 
maps appear nearly identical. Figure~\ref{fig:diffmap} shows the difference between the lower-left and lower-right
maps from Figure~\ref{fig:map}, on the same color scale, along with a simulated difference map 
for comparison. 
The feature at the location of one of the bright point sources is most likely caused by 
temporal variability of the source, as the SPT and \planck\ data were not taken simultaneously.

To make this result more quantitative, we could match the beam and filtering between these two 
data sets over the full SPT-SZ survey region, mask point sources, calculate the power in the difference
map, and compare that power to the expected power from noise alone. In the next 
section we do a nearly equivalent but somewhat simpler calculation. Using the cross-spectrum 
formalism outlined in Section~\ref{sec:powspec}, we calculate the \planck\ $143 \times 143$
power spectrum, the SPT $150 \times 150$ power spectrum, and the SPT-\planck\ $150 \times 143$
cross-spectrum, and we calculate the $\chi^2$ of the null hypothesis that these three spectra 
are measuring the same power. This calculation does not completely eliminate sample variance, 
as the perfectly mode-matched difference-map power spectrum would, but it strongly reduces
it (see Figure~\ref{fig:bandpower_error}).

\subsection{Bandpower Comparison and Recalibration}
\label{subsec:bp_comparison}

\begin{figure*}
\includegraphics[width=0.96\textwidth, trim=0 0.7cm 0 6cm]{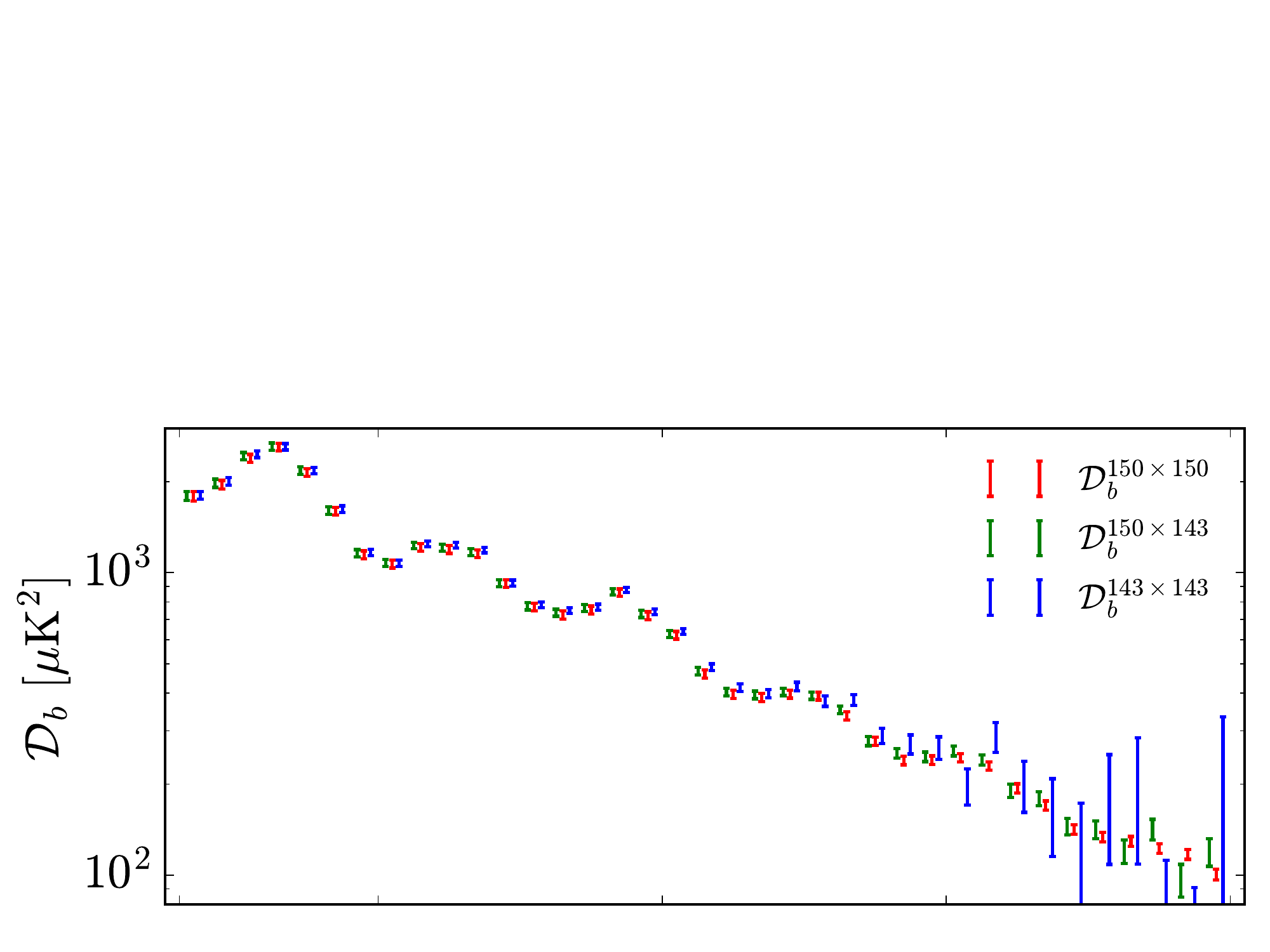}
\includegraphics[width=0.96\textwidth, trim=0 6.8cm 0 0.7cm]{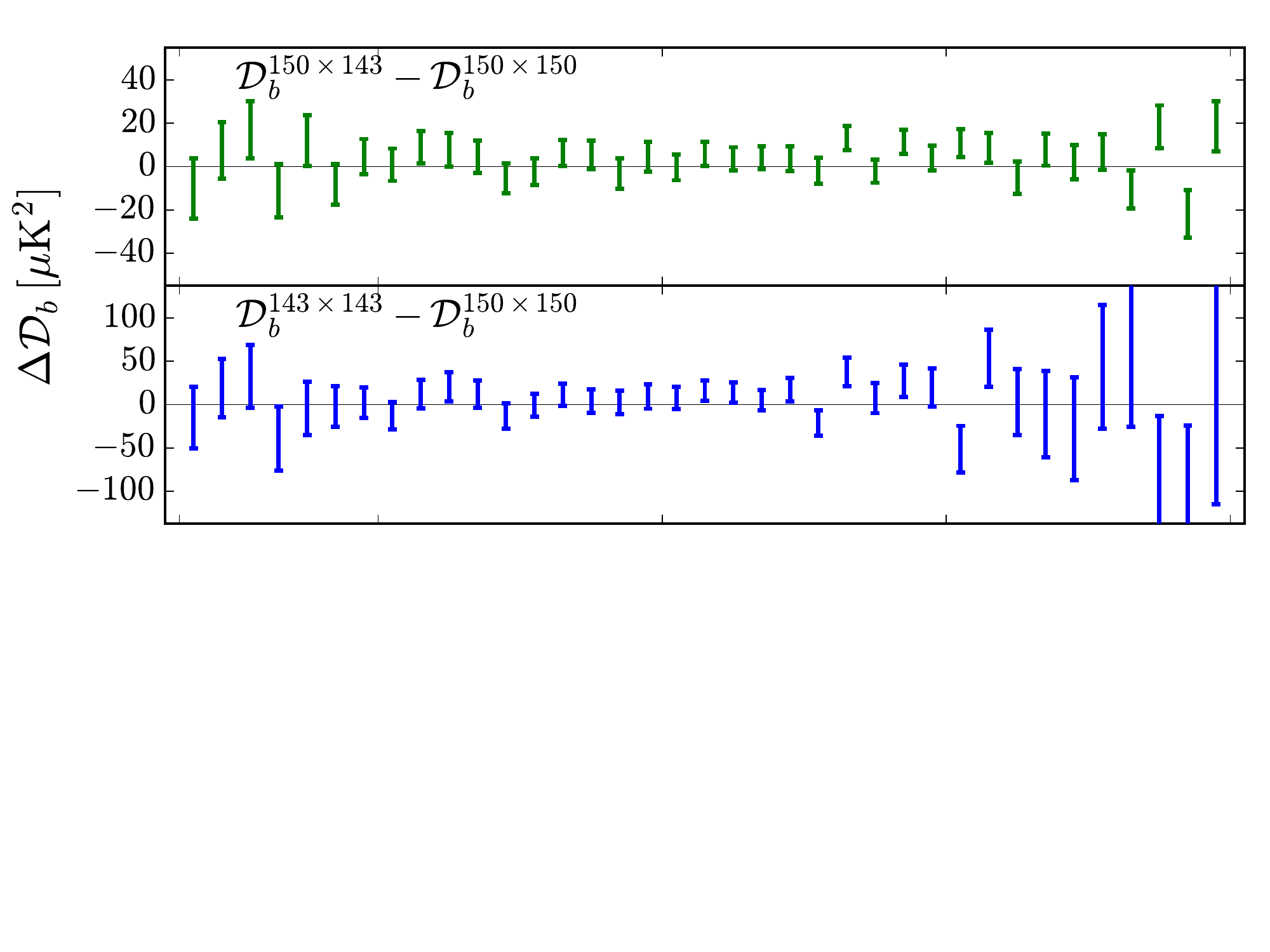}
\includegraphics[width=0.96\textwidth, trim=0 5cm 0 0.7cm]{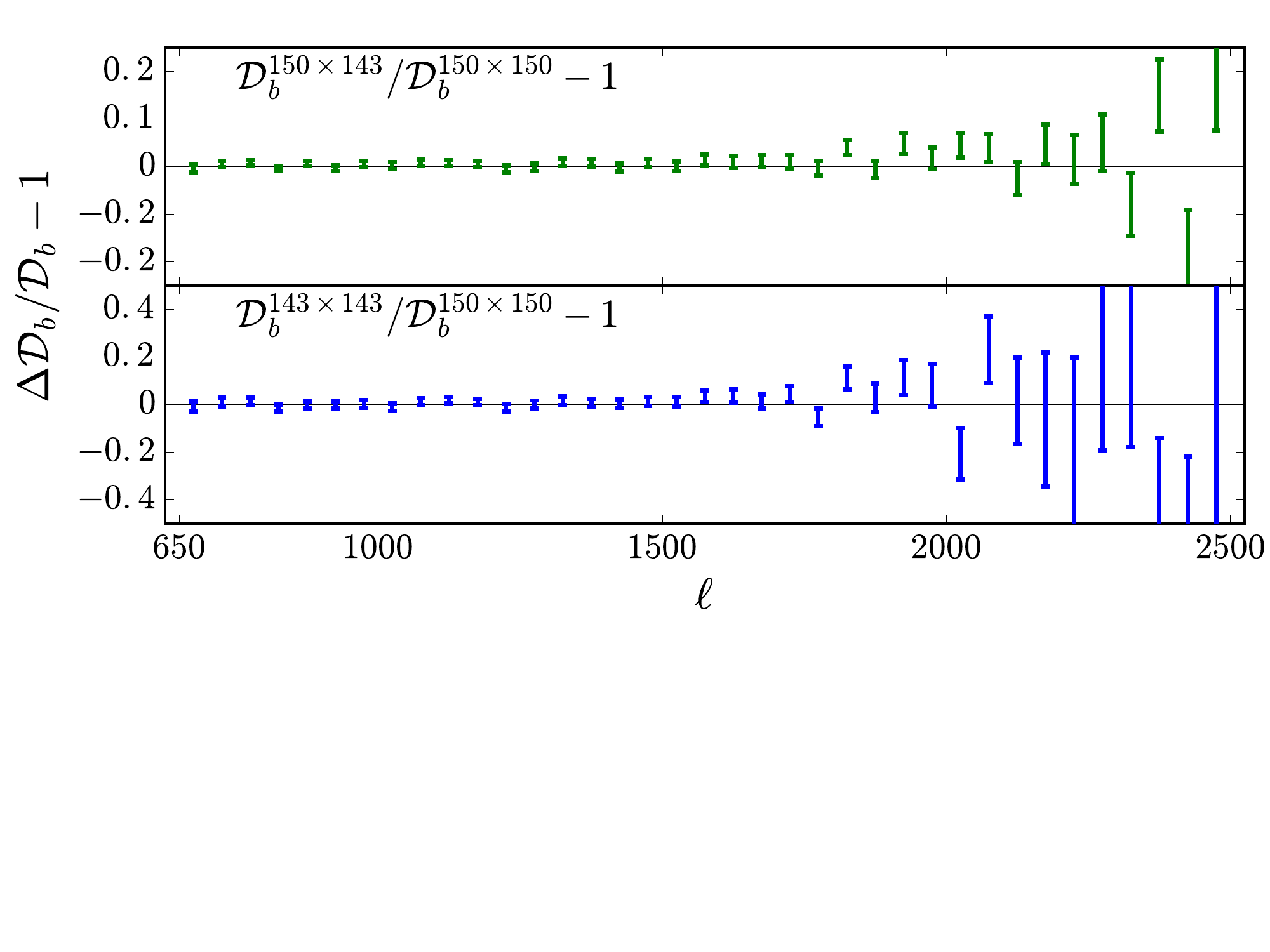}
\caption{\textit{Upper panel}: The window-function-corrected, unbiased bandpowers for $150\times 150$ (red), $150\times 143$ (green), and $143\times 143$ (blue). The error bars contain sample variance and noise variance obtained from our simulations, as well as beam uncertainties from the analytical calculations presented in Section~\ref{subsec:beam_error}. The green and blue error bars are offset horizontally for clarity. \textit{Middle panel}: The bandpower residuals $\mathcal{D}_b^{150\times 143} - \mathcal{D}_b^{150\times 150}$ (green) and $\mathcal{D}_b^{143\times 143} - \mathcal{D}_b^{150\times 150}$ (blue) with the window function correction. The error bars come from the diagonal of the covariance matrix of the bandpower residuals of Eq.~\ref{eqn:covmat}. Under our null hypothesis, the residuals are expected to be consistent with 0. Note the different plotting range between the two bandpower residuals. \textit{Lower panel}: The window-function-corrected bandpower ratios $\mathcal{D}_b^{150\times 143} / \mathcal{D}_b^{150\times 150} - 1$ (green) and $\mathcal{D}_b^{143\times 143} / \mathcal{D}_b^{150\times 150} - 1$ (blue). The best-fit recalibration parameter for SPT, $r_c$,  has been applied to all panels of this figure.}
\label{fig:residual}
\end{figure*}

In this section, we present the comparison between the three sets of window-function-corrected bandpowers,
which we denote as $\mathcal{D}_b^{150\times 150}$, $\mathcal{D}_b^{150\times 143}$, and $\mathcal{D}_b^{143\times 143}$. 
We first show all three sets of bandpowers in the upper panel of Figure~\ref{fig:residual}. The error bars contain contributions from sample variance, noise variance, and beam uncertainties.
By eye, the three sets of bandpowers look very consistent. At low $\ell$, the scatter among the three 
sets of points is much smaller than the errors on any one set; this is because the errors on all three are
dominated by sample variance in this regime, and sample variance is highly correlated among the
three data sets. This is also why the three sets of error bars are nearly the same size at low $\ell$ (though
the $143\times 143$ bandpowers have slightly smaller error bars in this $\ell$ range because the 
modes lost in the SPT TOD filtering process result in slightly increased sample variance).
Noise variance begins to dominate the $143\times 143$ error bars at $\ell \gtrsim 1700$ and the $150\times 143$ error bars at 
$\ell \gtrsim 2000$, while the $150\times 150$ error bars are sample-variance-dominated over the entire range plotted.

The remaining panels of Figure~\ref{fig:residual} show various comparisons among the three sets of bandpowers.
Two comparison schemes are applied: differences and ratios. The former is good for diagnosing additive effects, while the latter is more sensitive to multiplicative systematics. In both schemes, $\mathcal{D}_b^{150\times 150}$ is chosen as the fiducial bandpower set.
These comparison plots and the $\chi^2$ and the probability to exceed (PTE) statistics calculated below are thus testing the
following set of null hypotheses:
1) two sets of bandpower residuals
\begin{eqnarray*}
\langle \Delta\mathcal{D}_{b,c} \rangle & \equiv & \langle \mathcal{D}_b^{150\times 143} - \mathcal{D}_b^{150\times 150} \rangle = 0 \\
\langle \Delta\mathcal{D}_{b,a} \rangle & \equiv & \langle \mathcal{D}_b^{143\times 143} - \mathcal{D}_b^{150\times 150} \rangle = 0
\end{eqnarray*}
and 2) two sets of bandpower ratios 
\begin{eqnarray*}
\langle \Delta\mathcal{D}_{b,c} / \mathcal{D}_b^{150\times 150} \rangle & \equiv & \langle \mathcal{D}_b^{150\times 143} / \mathcal{D}_b^{150\times 150} \rangle -1 = 0 \\
\langle \Delta\mathcal{D}_{b,a} / \mathcal{D}_b^{150\times 150} \rangle & \equiv & \langle \mathcal{D}_b^{143\times 143} / \mathcal{D}_b^{150\times 150} \rangle - 1 = 0.
\end{eqnarray*}

The absolute calibration of the SPT data from \citet{george15} has a statistical uncertainty of 
$\sim$1\% at 150~GHz. We expect the bandpower comparison to be significantly more precise
than this, so before plotting the bandpower residuals and ratios and before testing the null hypotheses above, 
we apply a recalibration parameter $r_c$ to bandpowers containing SPT data:
\begin{eqnarray}
\label{eqn:residual_c}
\Delta\mathcal{D}_{b,c} & \equiv & r_c \mathcal{D}_b^{150\times 143} - r_c^2 \mathcal{D}_b^{150\times 150} \\
\Delta\mathcal{D}_{b,a} & \equiv & \mathcal{D}_b^{143\times 143} - r_c^2 \mathcal{D}_b^{150\times 150},
\label{eqn:residual_a}
\end{eqnarray}
and similarly for the bandpower ratios.
Because of the precision of the \citet{george15} calibration, we expect $r_c$ to be very close to 1.
We then calculate and minimize $\chi^2$ as a function of $r_c$:
\begin{equation}
\chi^2(r_c) = \Delta\mathbf{D}^T_\mathrm{b}(r_c) \Xi^{-1}\Delta\mathbf{D}_\mathrm{b}(r_c),
\end{equation}
and similarly for the bandpower ratios, where $\Delta\mathbf{D}_\mathrm{b}$ denotes the vector that includes both $\Delta\mathcal{D}_{b,c}$ and $\Delta\mathcal{D}_{b,a}$. We fit the two sets of
residuals or ratios simultaneously. In principle we should include the
recalibration parameter in an adjustment to the noise contribution to the covariance matrix, but we neglect it for
simplicity, with the justification that the correction is very small
at less than 1\%. In Figure~\ref{fig:residual}, we have included the best-fit recalibration parameter for SPT, $r_c$, in all three panels.

In the middle panel of Figure~\ref{fig:residual} we show the bandpower residuals $\Delta\mathcal{D}_{b,c}$ and $\Delta\mathcal{D}_{b,a}$ with error bars given by the square root of the diagonal elements of the full covariance matrix. This figure shows the residual bandpowers are consistent with zero given the errors. In the lower panels of the same figure we show the bandpower ratios. Similar to the residuals, the error bars of the bandpower ratios are the square root of the diagonal elements of the full covariance matrix. Qualitively, these results appear to be consistent with our null hypotheses. 

\begin{figure}
\includegraphics[width=0.54\textwidth, trim=1.8cm 0 0 0.5cm]{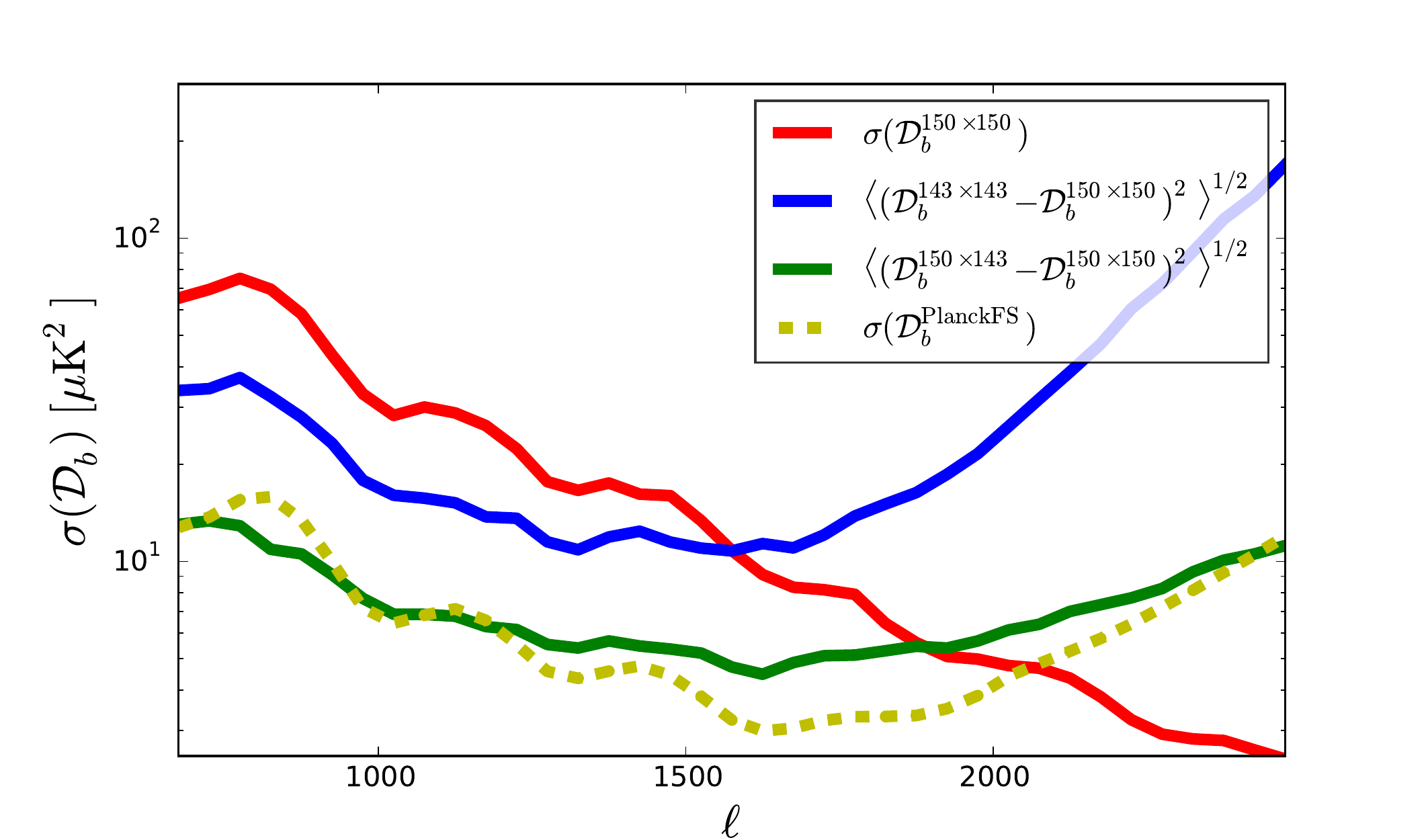}
\caption{Uncertainties in the SPT $150\times 150$ bandpowers (red) and in the two sets of bandpower residuals (not including beam uncertainty).
The sample variance is significantly reduced in the $150\times 150 - 143\times 143$ residuals
compared to the $150 \times 150$ bandpowers
(by roughly a factor of six in the lowest $\ell$ ranges, and by a larger factor
at higher $\ell$),
indicating that this comparison is more stringent than the comparison of $150\times 150$ to full-sky
\planck\ data, the uncertainty on which is dominated by the sample variance in the $150\times 150$ spectrum.
Also plotted are the bandpower uncertainties from full-sky \planck\ data, scaled to the $\ell$ bin size used in this work.
}
\label{fig:bandpower_error}
\end{figure}

\begin{table*}[!hbt]
\centering
\caption{Best-fit recalibration parameter, $\chi^2$, and PTE for the various bandpower comparisons.}
\begin{threeparttable}
 \begin{tabular}{ l c c c c}
 \hline\hline
Comparison & best $r_c$ & $\chi^2$ & PTE  \rule{0pt}{2.6ex} \\ [0.5ex]
 \hline
 {\bf Residual, combined, full covariance} & $1.0087 \pm0.0015$ & 78.7 & 0.30 \rule{0pt}{2.6ex} \\
  Residual, combined, no beam error & $1.0117 \pm0.0009$ & 89.8 & 0.09 \rule{0pt}{2.6ex} \\
 $150\times 143 - 150\times 150$, full covariance & $1.0087\pm 0.0016$ & 42.1 & 0.22 \rule{0pt}{2.6ex} \\
 $150\times 143 - 150\times 150$, no beam error & $1.0115\pm 0.0010$ & 50.4 & 0.06 \rule{0pt}{2.6ex} \\
 $143\times 143 - 150\times 150$, full covariance & $1.0076\pm 0.0022$ & 36.8 & 0.43 \rule{0pt}{2.6ex} \\
 $143\times 143 - 150\times 150$, no beam error & $1.0110\pm 0.0013$ & 41.8 & 0.23 \rule{0pt}{2.6ex} \\
 {\bf Ratio, combined, full covariance} & $1.0092 \pm 0.0015$ & 81.0 & 0.24 \rule{0pt}{2.6ex} \\
  Ratio, combined, no beam error & $1.0120\pm 0.0009$ & 91.0 & 0.08 \rule{0pt}{2.6ex} \\
 $(150\times 143) / (150\times 150)$, full covariance &  $1.0090\pm 0.0016$ & 43.3 & 0.19 \rule{0pt}{2.6ex} \\
 $(150\times 143) / (150\times 150)$, no beam error & $1.0118\pm 0.0010$ & 50.9 & 0.05 \rule{0pt}{2.6ex} \\
 $(143\times 143) / (150\times 150)$, full covariance & $1.0082\pm 0.0022$ & 38.1 & 0.37 \rule{0pt}{2.6ex} \\
 $(143\times 143) / (150\times 150)$, no beam error & $1.0113\pm 0.0013$ & 42.4 & 0.21 \rule{0pt}{2.6ex} \\
 \hline
\end{tabular}
\end{threeparttable}
\label{table:bpresults}
\end{table*}

The quantitative characterization comes from the $\chi^2$ value with the best-fit recalibration parameter. 
The best-fit values of $r_c$, minimum $\chi^2$, and probabilities to exceed that $\chi^2$ are listed
in Table~\ref{table:bpresults} for several combinations of data and covariance. The primary results
are from the combined residual bandpowers. Using the full $2\times 2$-block covariance matrix for the recalibration fit, these results give a best-fit of $r_c=1.0087\pm0.0015$ with $\chi^2 = 78.7$. There
are 37 $\ell$ bins in our analysis, so we have 74 data points among the two residual bandpowers and
one free parameter. The PTE for $\chi^2=78.7$ and 73 degrees of freedom is 0.30.
We find very similar results from the combined-ratio fit: $r_c=1.0092\pm0.0015$, $\chi^2 = 81.02$,
$\mathrm{PTE} = 0.24$.
Put another way, given the noise properties and the beam uncertainties of the two experiments, 30\% (24\%) of our simulations have a higher $\chi^2$ for the bandpower differences (ratios) than we find with the real data. 
We thus find as our primary result that the SPT and \planck\ data in the SPT-SZ sky patch are 
quite consistent with the null hypothesis that there is no systematic offset in the two experiments' measurement of the sky. A byproduct of this analysis is that we recalibrate the SPT data with a 
statistical precision of 0.30\% in power (0.15\% in temperature) relative to \planck\ 143 GHz.
This is comparable with the absolute calibration uncertainty of the \planck\ 143 GHz, 0.14\% in 
power (0.07\% in temperature, \citealt{planck15-8}), so we add this uncertainty in quadrature
for a final SPT 150~GHz calibration uncertainty of 0.33\% in power.

In Table~\ref{table:bpresults}, we also report the quantities of consistency from the single pair of bandpowers. 
For example, the PTE is 0.22 for the residual between $\mathcal{D}_b^{150\times 143}$ and $\mathcal{D}_b^{150\times 150}$ with
the full covariance matrix. For the other pair of residual bandpowers, 
$\mathcal{D}_b^{143\times 143}$ and $\mathcal{D}_b^{150\times 150}$,  the PTE is 0.43 with the full covariance matrix.
The results without the beam uncertainties are also listed in Table~\ref{table:bpresults}.

As discussed in Section~\ref{subsubsec:plancknoise}, the white-noise assumption in simulations for $\mathcal{D}_b^{143\times 143}$ results in an overestimate of the noise contribution to the covariance matrix, at
roughly the 10\% level in $\mathcal{D}_b$ errors, or the 20\% level in variance. 
If we assumed the \planck\ noise was the dominant contribution to the 
residual and ratio covariance in the $150\times 150$ vs.~$143\times 143$ comparisons,
we would expect roughly a 20\% increase in $\chi^2$ if we were able to use more realistic noise
simulations. The resulting $\chi^2$ would still correspond to a reasonable PTE ($\gtrsim 0.16$) for the null hypothesis.
This would also be true of the combined constraints, particularly because 
the main constraining power comes from $\mathcal{D}_b^{150\times 143}$, which is unaffected by the assumption of white noise in the \planck-only bandpowers.

Because of the largely reduced sample variance contribution to the difference between bandpowers within the same sky coverage (roughly a factor of six in $150\times 143 - 150\times 150$ compared to $150 \times 150$),
the comparison of $\mathcal{D}_b^{150\times 150}$ with the $\mathcal{D}_b^{150\times 143}$ and $\mathcal{D}_b^{143\times 143}$ bandpowers derived from within the SPT patch, provide tighter tests, over a wide range of angular scales, than can be achieved by the comparison of $\mathcal{D}_b^{150\times 150}$ with the more precise \planck\ spectra derived from nearly the full sky. In Figure~\ref{fig:bandpower_error}, we compare the uncertainties on the two sets of bandpower residuals $\mathcal{D}_b^{143\times 143} - \mathcal{D}_b^{150\times 150}$ (blue), and $\mathcal{D}_b^{150\times 143} - \mathcal{D}_b^{150\times 150}$ (green) to the uncertainties on the SPT-only bandpowers $\mathcal{D}_b^{150\times 150}$ (red).
 In the lower-$\ell$ region ($\ell<1800$), the green curve has the lowest error because the sample variance has been greatly reduced by subtracting the SPT bandpowers from the cross bandpowers, while at higher $\ell$ the green curve rises due to the \planck\ noise contribution. 
 In the most constraining case ($\mathcal{D}_b^{150\times 143} - \mathcal{D}_b^{150\times 150}$), the errors for almost all bins are $\le 10\mathrm{\mu K^2}$, comparable to the uncertainty in the 
PlanckFS bandpowers.
 Both sets of bandpower residuals yield very stringent tests on the consistency of the two datasets, and our results show that these datasets are formally consistent in this patch.

\subsection{Comparison to the Full-sky \planck\ 2015 TT High-$\ell$ 
Bandpowers}
\label{sec:tilt}

\begin{figure*}
\includegraphics[width=\textwidth, trim=3cm -1.5cm 4cm 0cm]{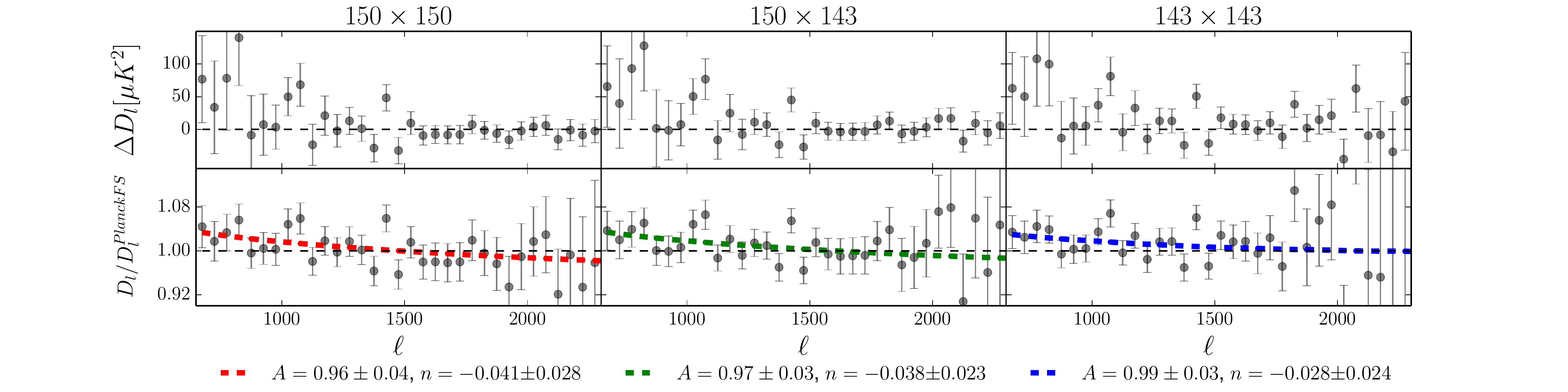}
\caption{\textit{Upper panel}: Residuals between the in-patch and PlanckFS bandpowers. \textit{Lower panel}: Ratios of the in-patch and PlanckFS bandpowers. The colored dashed lines are the best-fit power laws for each ratio.}
\label{fig:bpfsratio}
\end{figure*}

While restricting the bandpower comparison to overlapping sky substantially reduces differential sample variance, it does increase the \planck\ covariance. 
In this section, we instead investigate
differences between the ``in-patch'' bandpowers (i.e., the 150 $\times$ 150, 150 $\times$ 143, 
and 143 $\times$ 143 bandpowers from the 2540 deg$^2$ SPT-SZ survey region) and the 
full-sky \planck\ 2015 TT high-$\ell$ unbinned frequency combined bandpowers \citep{planck15-13}, which 
we refer to as ``PlanckFS.'' In this case, the null hypothesis is that any differences between
PlanckFS and the in-patch bandpowers are consistent with expectations given Gaussian
statistics and statistical isotropy. 

In Figure 9 we show the residuals and ratios for the in-patch to PlanckFS bandpowers. 
The residual plots show all three sets of the in-patch bandpowers prefer greater power at $650 < \ell < 1200$, indicating the \sptsz{} patch has greater power at these multipoles than the full-sky average. 
The ratio plots also suggest a tilt with respect to PlanckFS, although the significance is modest. 
The slope of the tilt is consistent to $0.5\,\sigma$ between the three cross-spectra, being slightly larger in $150\times 150$  and smaller in $143\times 143$. 

To quantify the statistical significance of this tilt we model the ratio of the in-patch and PlanckFS bandpowers as a power law:  
\be 
\frac{D^{th,i\times j}_{\ell}}{D^{\rm{PlanckFS}}_{\ell}}=A \left(\frac{\ell}{4000}\right)^n.
\ee 
We assume a Gaussian likelihood with 
\begin{eqnarray}
-2\ln\Lk(A,n|D^{i\times j},D^{\mathrm{PlanckFS}}) =  \Delta_b\Sigma_{bb'}^{-1}\Delta_b',
\\
\Delta_b = \frac{D^{i\times j}_b}{W^{i\times j}_{b\ell}D^{\mathrm{PlanckFS}}_\ell} - R_b^{i\times j}(A,n),
\\
 R_b^{i\times j}(A,n) = W^{i\times j}_{b\ell}\bigg(A\bigg(\frac{\ell}{4000}\bigg)^n + F^{i\times j}_{\ell}\bigg). 
\end{eqnarray}
Here $D^{\mathrm{PlanckFS}}_{\ell}$ is the best-fit PlanckFS power spectrum, 
$\Sigma_{bb'}=\langle \Delta_b\Delta_{b'}\rangle$ (only the SPT beam uncertainty is included and cross-correlations between $D^{i\times j}_b$ and the $ D^{\mathrm{PlanckFS}}_{\ell}$ are negligible), and $F^{i\times j}_{\ell}$ is the foreground model adopted from S13 with the frequency dependence from \citet{george15} included. 
There are three foreground amplitude parameters included in $F^{i\times j}_{\ell}$ to account for the SZ, Poisson-point-source, and clustered CIB uncertainties. 
The best-fit re-calibration from Section 4.3 has been applied to the in-patch SPT-only and
SPT-\planck\ bandpowers. 
The best-fit values for $A$ and $n$ are reported in Figure~\ref{fig:bpfsratio}.
We find only marginal evidence for a tilt between PlanckFS and any of
the three cross-spectra; the most significant tilt is for $150\times
150$ and is $1.5\,\sigma$ away from zero. We conclude that the tilts
we see in the observed spectra are roughly consistent with the expectation based on
Gaussian statistics and the assumption of statistical isotropy---i.e., they are roughly consistent with $n=0$.

\begin{table}[htb]
\centering
\caption{Power-law Parameter PTEs}
 \begin{tabular}{ c c c c c}
 \hline\hline
  & $\chi^2$ & PTE  \rule{0pt}{2.6ex} \\ [0.5ex]
 \hline
 $150\times 150 - 150\times 143$ & 0.33 & 0.41 \rule{0pt}{2.6ex} \\
 $150\times 150 - 143\times 143$ & 1.79 & 0.85\\
 $150\times 143 - 143\times 143$ & 1.18 & 0.55 \\ [1ex]
 \hline
\end{tabular}
\begin{tablenotes}
\small
\item The $\chi^{2}$ and PTE for the null hypothesis that the\\ 
tilt relative to PlanckFS is the same in the in-patch\\ 
$150\times150$, $150\times143$, and $143\times143$ bandpowers.\end{tablenotes}
\label{table:param PTE}
\end{table}

To determine the expected statistical properties of the
differences in the best-fit values of $A$ and $n$ we
construct a covariance matrix ($\vec{C}$) from the set of best-fit values of $A$ and $n$ in each of the 400 simulations of Section~\ref{subsec:sims}. 
We then calculate $\chi^2$ for these differences as
\begin{equation}
    \chi^2 = (\vec{\Theta}-\bar{\vec{\Theta}})^{\intercal}\vec{C}^{-1}(\vec{\Theta}-\bar{\vec{\Theta}})
\end{equation}
where $\vec{\Theta}$ is the vector of parameter differences and $\bar{\vec{\Theta}}$ is the mean simulation difference, which is consistent with zero. 
The breakdown of PTEs for the various pair differences is shown in Table~\ref{table:param PTE}. 
As can be seen, the most extreme PTE is 0.85 and the lowest is 0.41. We conclude that the observed tilts in the three cross-spectra are completely consistent with each other. 

All our tests are consistent with the following explanation
for the tilts we observe in the in-patch spectra relative to the best-fit \planck\ full-sky
spectrum: they are driven by a sample-variance fluctuation away from
the full-sky average, the magnitude of which is roughly consistent with
expectations under the assumption of statistical isotropy and our noise model.

\subsection{Pipeline Checks}
\label{sec:pipecheck}

All published power spectra from the SPT collaboration have been calculated using some
variant of the cross-spectrum pseudo-$C_\ell$ pipeline used in this work. Extensive checks
have been performed on this pipeline (see, e.g., Section 4.2 of \citealt{story13}), demonstrating
that the correct input spectrum is recovered from simulated data, even when that spectrum
differs from the spectrum assumed in calculating the filter transfer function. If, however, some
aspect of the pipeline were inducing a bias in the estimated power spectra (through some 
mechanism that has escaped all pipeline tests), this bias would affect both the SPT-SZ and
\planck\ data used in this work (because we have mock-observed the \planck\ data and 
analyzed it with the SPT pipeline). If the bias on the two data sets were comparable, it would then
divide or subtract out in the bandpower comparison, and we would (wrongly) conclude there was no
issue with either data set.

To test this scenario, we have analyzed the in-patch \planck\ data using an alternate pipeline.
Specifically, we created a HEALPix version of the SPT-SZ sky-patch and point-source
mask (stitched together from the individual-field masks), and we
handed this mask and the half-mission \planck\ 143 GHz maps to the
{\tt PolSpice}\footnote{\url{http://www2.iap.fr/users/hivon/software/PolSpice}}\citep{szapudi01,chon04}
code, which is designed to estimate the power spectra of masked full-sky maps and
properly account for the masking. We binned the $\Delta \ell=1$ {\tt PolSpice}
output into the $\Delta \ell=50$ bins used in the SPT pipeline using the
bandpower window functions calculated for the $143 \times 143$ ``scanned, filtered" \planck\
bandpowers. We then calculated the ratio of the ``unscanned, unfiltered" (PolSpice)
bandpowers to the scanned, filtered ones and found that they agree to better
than 3\% in every individual bin in which there is appreciable
signal-to-noise in the $143 \times 143$ spectrum, with an overall ratio of
$1.0028 \pm 0.0050$ over the range $600 < \ell < 1800$ and no evidence of a
trend with $\ell$. We have also re-done the tilt calculation in Section~\ref{sec:tilt} 
using the unscanned, unfiltered bandpowers and found results consistent
with what we found with the scanned, filtered bandpowers (within a fraction of a
sigma).  We thus conclude that our fundamental results are not an
artifact of the SPT analysis method.
 
\section{Conclusions}

In this paper, we have compared 150 GHz SPT data and 143 GHz \planck\ data in the same region of the sky,
namely the 2540 $\deg^2$ SPT-SZ survey footprint.
We have performed a visual comparison of maps constructed from the two data sets and found 
the difference between the two maps to be visually consistent with noise, once they have been 
filtered to display the same angular modes.

We then performed a quantitative analysis of the consistency of the maps, relying primarily on a comparison of the cross-spectrum of two halves of the SPT data with the SPT $\times$ \planck\ cross spectrum. We also compared the SPT $\times$ SPT spectrum with the cross-spectrum of two halves of the \planck\ data. 
These comparisons were made using differences between and ratios of two sets of binned power spectra,
or bandpowers, at a time, always using the SPT $\times$ SPT bandpowers as the fiducial set. To test
the null hypothesis that the bandpower differences (after recalibrating the SPT data) are consistent
with zero---or that the ratios are consistent with unity---we created a suite of 400 simulations of the signal 
and noise properties of the SPT and \planck\ maps, including signal contributions from the CMB and 
extragalactic foregrounds. 
We found our most stringent test, based on the expected variance of the differences, to be the comparison of the SPT $\times$ SPT spectrum with the SPT-\planck\ cross spectrum. Forming a $\chi^2$ quantity from these bandpower differences and the simulation-based covariance matrix, we have found a value that is exceeded by 22\% of the analogous $\chi^2$ values derived from the simulated data, i.e., corresponding to the PTE value 0.22. When we add the residuals between the SPT $\times$ SPT and \planck\ $\times$ \planck\ cross-spectra, we find a PTE of 30\%. All other tests result in similarly unremarkable PTEs. We find no evidence of a failure of our null model; i.e., we see no evidence for systematic errors or under- or over-estimate of statistical errors.

We have also compared the three sets of bandpowers from the 2540 deg$^2$ SPT-SZ survey region 
to the full-sky \planck\ 143~GHz power spectrum. 
Relative to the \planck\ full-sky spectrum, we have found a hint for a tilt in the in-patch bandpowers. 
For all three sets of in-patch bandpowers, the amplitude of the tilts we have obtained are consistent with each other and roughly consistent with expected noise and sample variance fluctuations.

This work shows that the SPT 150 GHz and \planck\ 143 GHz data are in very good agreement with each other within the 2540 deg$^2$ SPT-SZ survey area. In a companion paper \citep{aylor17}, we extend this comparison to the cosmological parameters that can be derived from these bandpowers. 

\acknowledgements{The South Pole Telescope is supported by the National Science Foundation through grant PLR-1248097.  Partial support is also provided by the NSF Physics 
Frontier Center grant PHY-1125897 to the Kavli Institute of Cosmological Physics at the University of Chicago, the Kavli Foundation and the Gordon and 
Betty Moore Foundation grant GBMF 947. The McGill group acknowledges funding from the National Sciences and Engineering Research Council of Canada, Canada Research Chairs program, and the Canadian Institute for Advanced Research.
Argonne National Laboratory work was supported under U.S. Department of Energy contract DE-AC02-06CH11357. 
BB is supported by the Fermi Research Alliance, LLC under Contract No. De-AC02-07CH11359 with the United States Department of Energy. 
CR acknowledges support from a Australian Research CouncilÕs Future Fellowship (FT150100074). We thank D. Rapetti for comments on the manuscript, 
and we thank an anonymous referee for helpful suggestions.

\begin{appendix}

In this appendix, we explain  why we use a different power spectrum estimator for the SPT bandpowers than in S13, and  discuss the potential difference this makes to our comparisons. 
S13 averaged the cross-spectra between $\mathcal{O}(100)$ single-observation maps, while in this work we use the cross-spectrum between two half-survey maps in each field. 
Similarly, S13 estimated the bandpower covariance matrix from the distribution of said cross-spectra, while we estimate the covariance from signal+noise simulations. 

The decision to change estimators is driven by the desire to use the same procedure for the $150\times 150$,  $150\times 143$, and $143\times 143$ bandpowers. 
With only the two half-survey maps for 143 GHz, the S13 covariance estimator would not work. 
Instead, we rely on signal+noise simulations to calculate the bandpower covariance. 
However, we could only create one noise realization per map for the single-observation maps used in S13. 
By using half-survey maps (i.e.~by coadding many single-observation maps), we can increase the number of independent noise realizations dramatically. 
As laid out in \S\ref{subsec:sptnoise}, we generate noise realizations by first nulling the signal in single-observation maps by differencing the left-going and right-going scans, and then coadding these noise maps with a random +1 or -1 prefactor. 
These noise realizations can then be added to the simulated signal-only maps to yield robust signal+noise SPT map simulations.

One might worry that the change in estimator could affect our comparison. 
To address this, we perform the following quantitative test. 
We replace the SPT 150 GHz half1-half2 cross bandpower by the original S13 bandpowers and redo the test on the bandpower residuals (with the covariance matrix of the residual unchanged). 
We find no significant differences. 
For $\mathcal{D}^{150\times 143 - 150\times 150}_b$ only with the S13 bandpowers, the best-fit $r_c = 1.0090\pm0.0016$ with $\chi^2 = 47.21$ and $\mathrm{PTE} = 0.100$. 
Recall from Table~\ref{table:bpresults}, the numbers with the half-survey bandpowers are  $r_c = 1.0087\pm0.0016$ with $\chi^2 = 42.07$ and $\mathrm{PTE} = 0.224$. 
For $\mathcal{D}^{143\times 143 - 150\times 150}_b$ with the S13 bandpowers,  the best-fit $r_c = 1.0086\pm 0.0022$ with $\chi^2=39.17$ and $\mathrm{PTE} = 0.330$.
Again from Table~\ref{table:bpresults}, the equivalent numbers with the half-survey bandpowers are  $r_c = 1.0076\pm0.0022$ with $\chi^2 = 36.83$ and $\mathrm{PTE} = 0.430$.
We therefore conclude that the change in SPT power spectrum estimators does not significantly impact our comparison of the SPT 150 GHz map and Planck 143 GHz map over the same region of sky.

\end{appendix}

\bibliography{../../../BIBTEX/spt}

\end{document}